\documentclass[12pt]{article}
\usepackage{amsfonts}
\usepackage{amssymb}
\usepackage{amsbsy}
\usepackage{mathrsfs}
\usepackage{amsmath}
\usepackage{graphicx}    
\usepackage{rotating}    
\usepackage{epsfig}
\usepackage{color}

\input epsf.sty

\newlength{\TZ}
\newcommand{\BEQ}{\begin{equation}}     
\newcommand{\BEA}{\begin{eqnarray}}
\newcommand{\BD}{\begin{displaymath}}
\newcommand{\EEQ}{\end{equation}}       
\newcommand{\EEA}{\end{eqnarray}}
\newcommand{\ED}{\end{displaymath}}
\newcommand{\vep}{\varepsilon}          
\newcommand{\vph}{\varphi}              
\newcommand{\vro}{\varrho}              
\newcommand{\D}{{\rm d}}                
\newcommand{\II}{{\rm i}}               
\newcommand{\sign}{{\rm sign\,}}        
\newcommand{\demi}{\frac{1}{2}}         
\newcommand{\wit}[1]{\widetilde{#1}}    
\newcommand{\wht}[1]{\widehat{#1}}      

\renewcommand{\vec}[1]{\boldsymbol{#1}} 

\definecolor{gruen}{rgb}{0,0.6,0}      
\definecolor{rot}{rgb}{0.75,0,0}        

\begin{document}

\begin{titlepage}

\vskip 1.5 cm
\begin{center}
{\Large \bf Non-local meta-conformal invariance, \\ diffusion-limited erosion and the XXZ chain}
\end{center}

\vskip 2.0 cm
\centerline{{\bf Malte Henkel}$^{a,b}$} 
\vskip 0.5 cm
\begin{center}
$^a$Rechnergest\"utzte Physik der Werkstoffe, Institut f\"ur Baustoffe (IfB), \\ ETH Z\"urich, Stefano-Franscini-Platz 3, 
CH - 8093 Z\"urich, Switzerland\\~\\
$^b$Groupe de Physique Statistique, D\'epartement de Physique de la Mati\`ere et des Mat\'eriaux, 
Institut Jean Lamour {\small (CNRS UMR 7198)}, Universit\'e de Lorraine Nancy,  \\
B.P. 70239, F - 54506 Vand{\oe}uvre-l\`es-Nancy Cedex, France\footnote{permanent address}
\end{center}

\begin{abstract}
Diffusion-limited erosion is a distinct universality class of fluctuating interfaces. 
Although its dynamical exponent $z=1$, none of the known variants of conformal invariance can act as its dynamical symmetry. 
In $d=1$ spatial dimensions, its infinite-dimensional dynamic symmetry is constructed and shown to be isomorphic to 
the direct sum of three loop-Virasoro algebras, with the maximal finite-dimensional
sub-algebra $\mathfrak{sl}(2,\mathbb{R})\oplus\mathfrak{sl}(2,\mathbb{R})\oplus\mathfrak{sl}(2,\mathbb{R})$. 
The infinitesimal generators are spatially non-local and use the Riesz-Feller fractional derivative. 
Co-variant two-time response functions are derived and reproduce the exact solution of diffusion-limited erosion. 
The relationship with the terrace-step-kind model of vicinal surfaces and the integrable XXZ chain are discussed.
\end{abstract}

\vfill

\end{titlepage}

\setcounter{footnote}{0}

\section{Introduction}

Symmetries have since a long time played an important r\^ole in the analysis of physical systems. The insight gained can be either
calculational, in that a recognised symmetry becomes useful in simplifying calculations, or else conceptual, in that the identification
of symmetries can lead to new level of understanding. In the statistical physics of equilibrium second-order phase transitions in two dimensions, 
{\em conformal invariance} has ever since the pioneering work of Belavin, Polyakov and Zamolodchikov \cite{Belavin84} created considerable progress, 
both computationally as well as conceptually. It then appears natural to ask if one might find extensions of conformal invariance which apply
to time-dependent phenomena. Here, we shall inquire about dynamical symmetries of the following stochastic Langevin equation, 
\textcolor{black}{to be called {\em diffusion-limited erosion ({\sc dle}) Langevin equation}, which reads} in momentum space  
\BEQ \label{1}
\D\wht{h}(t,\vec{q}) = -\nu |\vec{q}| \wht{h}(t,\vec{q})\D t + \wht{\jmath}(t,\vec{q})\D t + \left( 2\nu T\right)^{1/2}\D\wht{B}(t,\vec{q})
\EEQ
\textcolor{black}{and describes the Fourier-transformed height 
$\wht{h}(t,\vec{q})=(2\pi)^{-d/2}\int_{\mathbb{R}^d} \!\D \vec{r}\: e^{-\II\vec{q}\cdot\vec{r}} h(t,\vec{r})$. 
Because of the (Fourier-transformed) standard brownian motion $\wht{B}$, with the variance 
$\langle \wht{B}(t,\vec{q}) \wht{B}(t',\vec{q}')\rangle = \min(t,t')\delta(\vec{q}+\vec{q}')$, this is a stochastic process, called
{\em diffusion-limited erosion ({\sc dle}) process}. Herein, $\nu,T$ are non-negative constants and
$\delta(\vec{q})$ is the Dirac distribution. Since we shall be interested in deriving linear responses, an external infinitesimal source term
$\wht{\jmath}(t,\vec{q})$ is also included, to be set to zero at the end.}  
Inverting the Fourier transform in order to return to direct space, eq.~(\ref{1}) implies spatially long-range interactions. The conformal
invariance of equilibrium critical systems with long-range interactions has been analysed recently \cite{Paulos16}. 
Eq.~(\ref{1}) arises in several distinct physical contexts. 

\begin{figure}[tb]
\centerline{\includegraphics[scale=0.75]{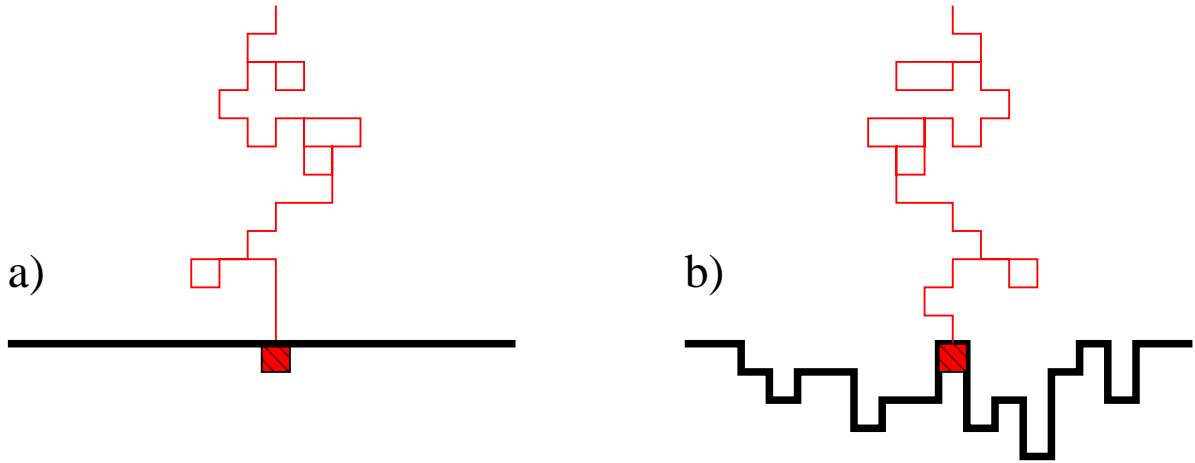}} 
\caption[fig1]{Schematics of the genesis of an eroding surface through the {\sc dle} process. 
(a) Initial state: a diffusing particle (red path) arrives on a flat surface (full black line) and erodes a small part of it. 
(b) Analogous process at a later time, when the surface has been partially eroded.  
\label{fig1}}
\end{figure}

\noindent
{\bf Example 1:} 
\textcolor{black}{For the original definition of the ({\sc dle}) process} \cite{Krug81}, one considers how an initially flat interface is 
affected by the diffusive motion of corrosive particles. A single corrosive particle starts initially far away from
the interface. After having undergone diffusive motion until the particle 
finally arrives at the interface, it erodes a particle from that interface.
Repeating this process many times, an eroding interface forms which 
is described in terms of a fluctuating height $h(t,\vec{r})$, see figure~\ref{fig1}. 
It can be shown that this leads to the {\sc dle} Langevin equation (\ref{1}) \cite{Krug81,Krug94}. 

\noindent Several lattice formulations of the model \cite{Krug81,Yoon03,Aarao07,Zoia07} confirm the dynamical exponent $z=1$. 

\newpage 

\noindent
{\bf Example 2:} 
A different physical realisation of eq.~(\ref{1}) invokes vicinal surfaces. 
Remarkably, for $d=1$ space dimension, the Langevin equation (\ref{1}) has been argued \cite{Spohn99} to be related to a system of 
non-interacting fermions, conditioned to an a-typically large flux. 
Consider the {\bf terrace-step-kink model} of a vicinal surface, and interpret 
the steps as the world lines of fermions. Its transfer matrix 
is the matrix exponential of the quantum hamiltonian $H$ of the asymmetric XXZ chain \cite{Spohn99}. 
Use Pauli matrices
$\sigma_{n}^{\pm,z}$, attached to each site $n$, such that the particle number at each site is $\vro_n=\demi\left(1+\sigma_{n}^{z}\right)=0,1$.
On a chain of $N$ sites, consider the quantum hamiltonian \cite{Spohn99,Popk11,Karevski16}
\BEQ \label{xxz}
H = - \frac{w}{2} \sum_{n=1}^{N} \left[ 2 v \sigma_n^{+}\sigma_{n+1}^{-} + 2 v^{-1} \sigma_{n}^{-}\sigma_{n+1}^{+} 
+\Delta \left( \sigma_{n}^{z}\sigma_{n+1}^{z}-1\right) \right]
\EEQ
where $w=\sqrt{pq\,}\, e^{\mu}$, $v=\sqrt{p/q\,}\,e^{\lambda}$ and 
$\Delta=2\left(\sqrt{p/q}+\sqrt{q/p}\right)e^{-\mu}$. Herein, $p,q$ describe the left/right bias of single-particle hopping and $\lambda,\mu$ are the
grand-canonical parameters conjugate to the current and the mean particle number. 
In the continuum limit, the particle density $\vro_n(t) \to \vro(t,r)=\partial_r h(t,r)$ 
is related to the height $h$ which in turn obeys (\ref{1}), with a {\em gaussian white noise} $\eta$ \cite{Spohn99}. This follows from the
application of the theory of fluctuating hydrodynamics, see \cite{Spohn14,Bert15} for recent reviews. 
The low-energy behaviour of $H$ yields the dynamical exponent $z=1$ \cite{Spohn99,Popk11,Karevski16}.
If one conditions the system to an a-typically large current, 
the large-time, large-distance behaviour of (\ref{xxz}) has very recently been
shown \cite{Karevski16} (i) to be described by a conformal field-theory with central charge $c=1$ and (ii) 
the time-space scaling behaviour of the stationary structure function has been worked out explicitly, for $\lambda\to\infty$. 
Therefore, one may conjecture that the so simple-looking eq.~(\ref{1}) should furnish an effective continuum
description of the large-time, long-range properties of quite non-trivial systems, such as (\ref{xxz}). 

\begin{figure}[tb]
\includegraphics[scale=0.4]{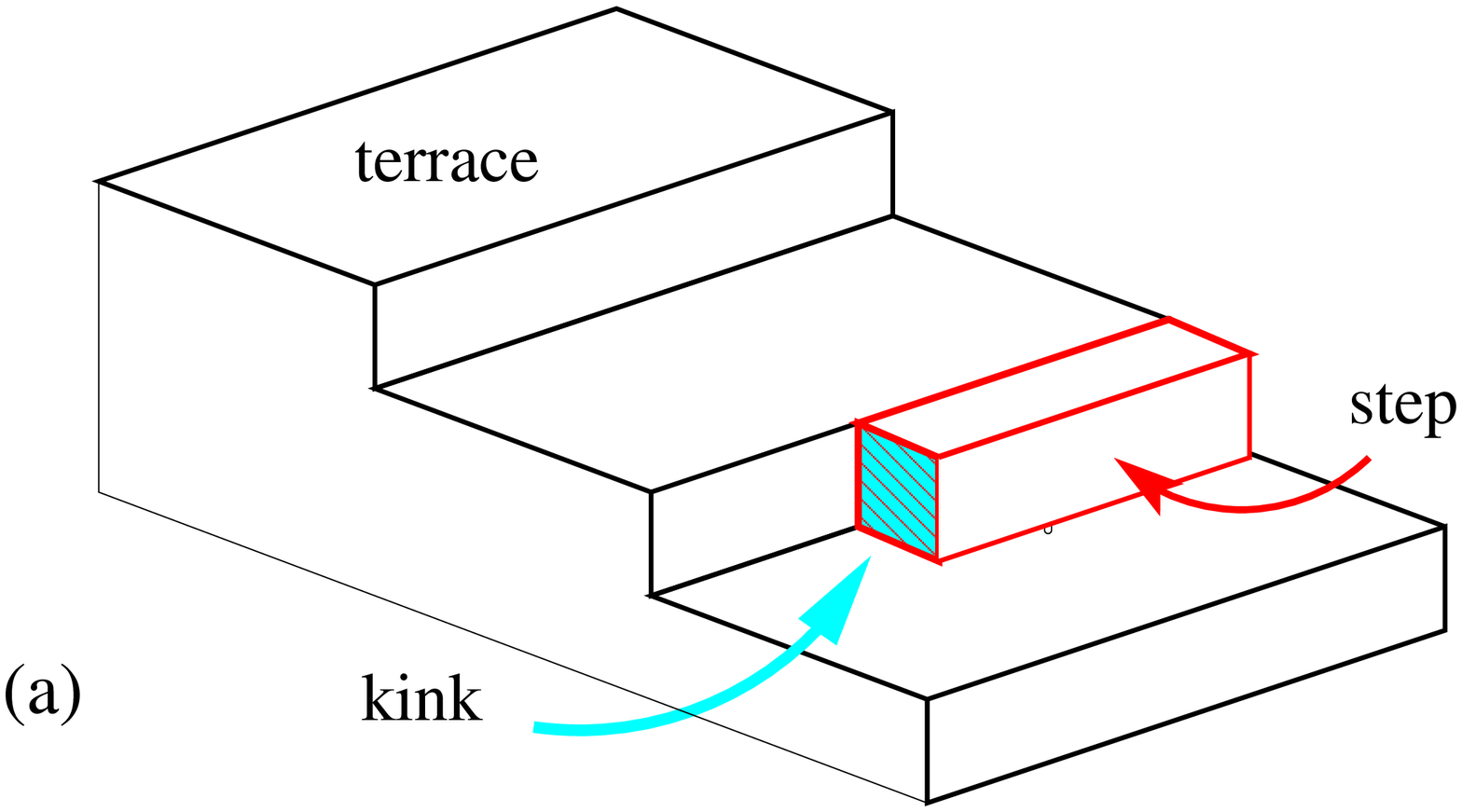}~~~~~~~\includegraphics[scale=0.4]{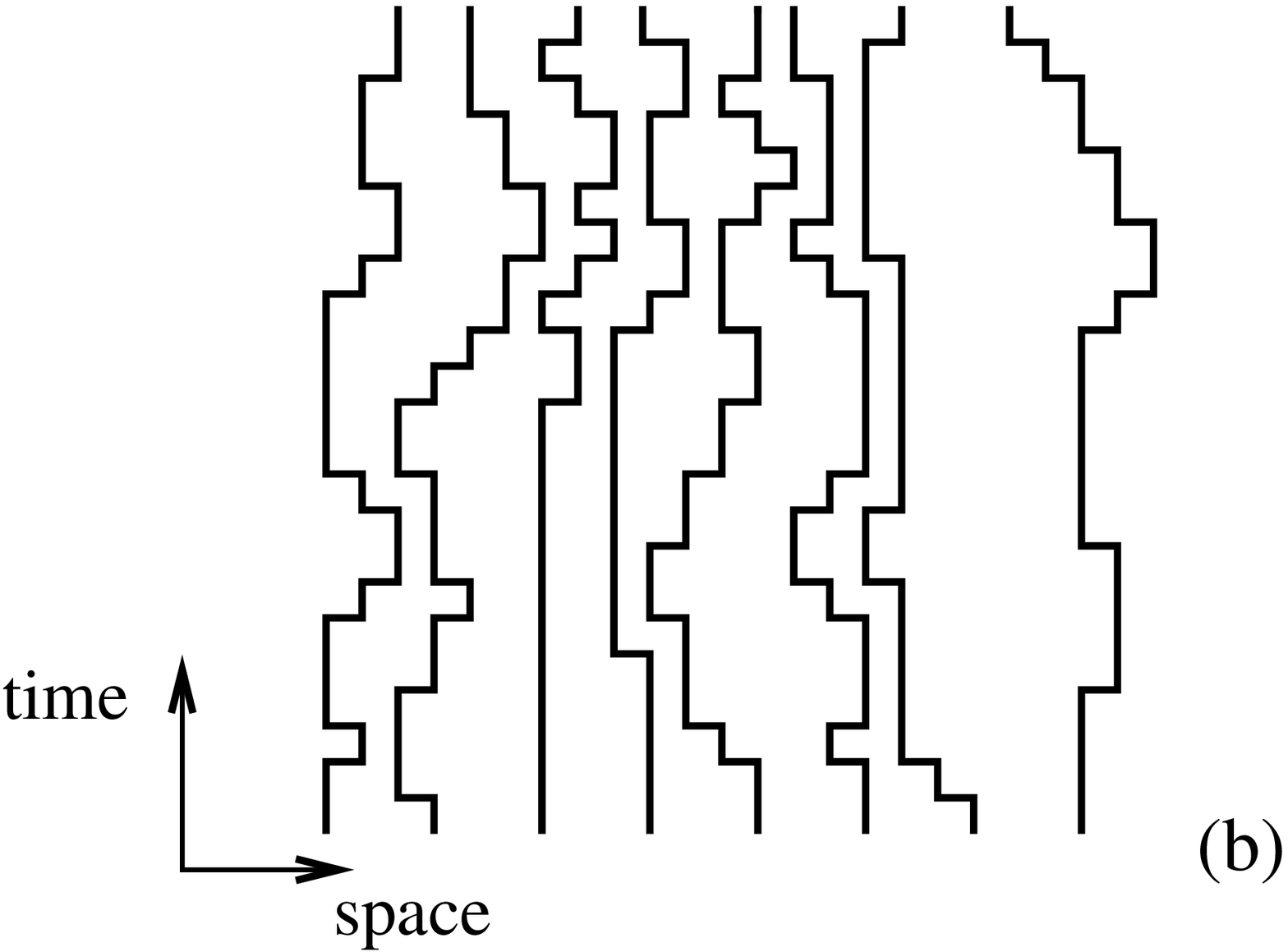} 
\caption[fig2]{(a) Schematic illustration of a vicinal surface, formed by terraces. 
Fluctuations between terrasses are described by steps and kinks. (b) Reinterpretation of the steps of a 
vicinal surface as non-intersecting world lines in $1+1$ dimensions of an ensemble of fermionic particles. 
\label{fig2}}
\end{figure}

The physical realisation of eq.~(\ref{1}) in terms of the {\sc dle} process makes it convenient to discuss the results in terms
of the  physics of the growth of interfaces \cite{Barabasi95,Halpin95,Krug97}, which can be viewed as a paradigmatic example of 
the emergence of non-equilibrium collective phenomena \cite{Henkel10,Taeuber14}. Such an interface can be described in terms of a time-space-dependent
height profile $h(t,\vec{r})$. This profile depends also on the eventual fluctuations of the set of initial states and on the noise
in the Langevin equation, hence $h$ should be considered as a random variable. The degree of fluctuations can be measured through the
interface width. If the model is formulated first on a hyper-cubic lattice $\mathscr{L}\subset\mathbb{Z}^d$ of $|\mathscr{L}|=L^d$ sites, 
the interface width is defined by 
\BEQ \label{2}
w^2(t;L) := \frac{1}{L^d} \sum_{\vec{r}\in\mathscr{L}} \left\langle \left( h(t,\vec{r})-\overline{h}(t) \right) \right\rangle^2 
= L^{2\beta z} f_w\left(t L^{-z}\right) 
\sim \left\{ \begin{array}{ll} t^{2\beta} & \mbox{\rm ~;~~ if $tL^{-z}\ll 1$} \\
                              L^{2\beta z} & \mbox{\rm ~;~~ if $tL^{-z}\gg 1$} \end{array} \right.
\EEQ
where the generically expected scaling form, for large times/lattice sizes $t\to\infty$, $L\to\infty$, is also indicated. 
Physicists call this {\bf Family-Vicsek scaling} 
\cite{Family85}.\footnote{Implicitly, it is assumed here that one is {\em not} at the `upper critical dimension $d^*$', 
where this power-law scaling is replaced by a logarithmic scaling form, see also below.} 
Herein, $\langle .\rangle$ denotes an average over many independent samples and 
$\overline{h}(t) := L^{-d} \sum_{\vec{r}\in\mathscr{L}} h(t,\vec{r})$ is the spatially averaged height. Furthermore, 
$\beta$ is called the {\bf growth exponent}, $z>0$ is the {\bf dynamical exponent} and $\alpha:=\beta z$ is 
the {\bf roughness exponent}. 
When $t L^{-z}\gg 1$, one speaks of the {\bf saturation regime} and when $t L^{-z}\ll 1$, one speaks of the {\bf growth regime}. 
We shall focus on the growth regime from now on. 

\noindent {\bf Definition 1:}
{\it On a spatially infinite substrate, an interface with a width $w(t)\nearrow\infty$ for large times $t\to\infty$ is called {\bf rough}. 
If $\lim_{t\to\infty} w(t)$ is finite, the interface is called {\bf smooth}.}

\noindent 
A first appreciation of the nature of the interface follows: if in (\ref{2}) $\beta>0$, the interface is rough. 

In addition, dynamical properties of the interface can be studied through the two-time correlators and responses. 
In the growth regime (where effectively $L\to\infty$),
one considers the double scaling limit $t,s\to\infty$ with $y :=t/s>1$ fixed and expects the scaling behaviour
\begin{subequations}  \label{4} 
\begin{align}
C(t,s;\vec{r}) &:= \left\langle \left(h(t,\vec{r}) - \left\langle \overline{h}(t)\right\rangle \right)
\left(h(s,\vec{0}) - \left\langle \overline{h}(s)\right\rangle \right) \right\rangle 
\hspace{0.55truecm}\:=\: s^{-b} F_C\left( \frac{t}{s}; \frac{\vec{r}}{s^{1/z}} \right)  \label{4.C} \\
R(t,s;\vec{r}) &:= \left. \frac{ \delta \left\langle h(t,\vec{r}) - \overline{h}(t)\right\rangle}{\delta j(s,\vec{0})}\right|_{j=0} 
\:=\: \left\langle h(t,\vec{r}) \wit{h}(s,\vec{0}) \right\rangle \:=\: s^{-1-a} F_R\left( \frac{t}{s}; \frac{\vec{r}}{s^{1/z}} \right)
\label{4.R}
\end{align}
\end{subequations}
where $j$ is an external field conjugate\footnote{In the context of Janssen-de Dominicis theory, $\wit{h}$ is the conjugate response field to $h$, 
see \cite{Taeuber14}.\textcolor{black}{Throughout, all correlators are calculated with $j=0$.}} to the height $h$. 
Spatial translation-invariance was implicitly admitted in (\ref{4}). This defines the {\em ageing exponents} $a,b$. 
The {\bf autocorrelation exponent} $\lambda_C$ and the {\bf autoresponse exponent} $\lambda_R$ are defined from the asymptotics
$F_{C}(y,\vec{0}) \sim y^{-\lambda_{C}/z}$ and $F_{R}(y,\vec{0}) \sim y^{-\lambda_{R}/z}$, respectively, as $y\to\infty$. 
For these non-equilibrium exponents, one has $b=-2\beta$ \cite{Krug97} 
and the bound $\lambda_C\geq (d+z b)/2$ \cite{Yeun96,Henkel15b}.

For the {\sc dle} process, these exponents are readily found form the exact solution of (\ref{1}) \cite{Krug81,Krug94,Henkel16a}. 
For an initially flat interface $h(0,\vec{r})=0$, the two-time correlator and response are in Fourier space
\begin{subequations} \label{6}
\begin{align}
\wht{C}(t,s;\vec{q},\vec{q}') &:= \left\langle \wht{h}(t,\vec{q}) \wht{h}(s,\vec{q}') \right\rangle
= \frac{T}{|\vec{q}|} \left[ e^{-\nu|\vec{q}||t-s|} - e^{-\nu|\vec{q}|(t+s)} \right] \delta(\vec{q}+\vec{q}'),
\label{6C} \\
\wht{R}(t,s;\vec{q},\vec{q}') &:= \left. \frac{\delta \langle \wht{h}(t,\vec{q})\rangle}{\delta \wht{\jmath}(s,\vec{q}')}\right|_{j=0} 
\hspace{0.4truecm}= \Theta(t-s)\; e^{-\nu|\vec{q}|(t-s)}\, \delta(\vec{q}+\vec{q}').
\label{6R}
\end{align}
\end{subequations}
In direct space, this becomes, for $d\ne 1$ and with ${\cal C}_0 := \Gamma((d+1)/2)/(\Gamma(d/2)\pi^{(d+1)/2})$
\begin{subequations} \label{7}
\begin{align}
C(t,s;\vec{r}) &= \frac{T {\cal C}_0}{d-1} 
\left[ \left( \nu^2 (t-s)^2 + r^2 \right)^{-(d-1)/2} - \left( \nu^2 (t+s)^2 + r^2 \right)^{-(d-1)/2} \right]
\label{7C} \\
R(t,s;\vec{r}) &= {\cal C}_0\: \Theta(t-s)\:\nu(t-s) \left(\nu^2 (t-s)^2 + r^2\right)^{-(d+1)/2} 
\label{7R}
\end{align}
\end{subequations}
where the Heaviside function $\Theta$ expresses the causality condition $t>s$. In particular, in the growth regime, the interface width 
reads\footnote{Technically, a high-momentum cut-off $\Lambda$ for $d>1$ is needed in the $L\to\infty$ limit. ${\cal C}_1(\Lambda)$ is a known constant.}
\BEQ \label{8}
\!\!w^2(t) = C(t,t;\vec{0}) = \frac{T {\cal C}_0}{1-d} \left[ \left( 2\nu t\right)^{1-d} - {\cal C}_1(\Lambda) \right] \stackrel{t\to\infty}{\simeq}
\left\{ \begin{array}{ll} \!\!T{\cal C}_0{\cal C}_1(\Lambda)/(d-1)        & \mbox{\rm ;\, if $d>1$} \\
                          \!\!T{\cal C}_0 \ln( 2\nu t)                    & \mbox{\rm ;\, if $d=1$} \\
                          \!\!T{\cal C}_0 (2\nu)^{1-d}/(1-d)\cdot t^{1-d} & \mbox{\rm ;\, if $d<1$}
        \end{array} \right.
\EEQ
Hence $d^*=1$ is the upper critical dimension of {\sc dle} process. It follows that at late times the {\sc dle}-interface is 
smooth for $d>1$ and rough for $d\leq 1$. 
On the other hand, one may consider the {\it stationary} limit $t,s\to\infty$ with the time difference 
$\tau=t-s$ being kept fixed. Then one finds a fluctuation-dissipation relation
$\partial C(s+\tau,s;\vec{r})/\partial \tau = - \nu T R(s+\tau,s;\vec{r})$. 
The similarity of this to what is found for equilibrium systems is unsurprising, since several discrete lattice variants of the {\sc dle} process exist  
and are formulated as an equilibrium system \cite{Yoon03}. Lastly, the exponents defined above are read off by taking the scaling limit, 
and are listed in table~\ref{tab1}. 
In contrast to the interface width $w(t)$, which shows a logarithmic growth at $d=d^*=1$, logarithms cancel in the two-time correlator $C$ and
response $R$, up to {\it additive} logarithmic corrections to scaling. 
This is well-known in the physical ageing at $d=d^*$ of simple magnets or of the Arcetri model \cite{Hase06,Ebbinghaus08,Henkel15b}.

\begin{sidewaystable}
\caption[tab1]{Exponents of growing interfaces in the Kardar-Parisi-Zhang ({\sc kpz}), 
Edwards-Wilkinson ({\sc ew}), Arcetri (for both $T=T_c$ and $T<T_c$) and {\sc dle} universality classes.   
The numbers in bracket give the estimated error in the last digit(s).  \label{tab1}} 
\begin{center}\begin{tabular}{||l|ccccccc|l||} \hline\hline
model       & ~$d$~ & ~$z$~        & ~$\beta$~      & ~$a$~     & ~$b$~          & ~$\lambda_C$~ & ~$\lambda_R$~ & Ref. \\ \hline\hline
{\sc kpz}   & $1$   & $3/2$        & $1/3$          & $-1/3$    & $-2/3$         & $1$           & $1$       & \cite{Kardar86,Krech97,Henkel12} \\
            & $2$   & $1.61(2)$~~  & $0.2415(15)$   & $0.30(1)$ & $-0.483(3)$~~  & $1.97(3)$     & $2.04(3)$ & \cite{Odor14} \\
            & $2$   & $1.61(2)$~~  & $0.241(1)$~~~~ &           & $-0.483$~~~~~~ & $1.91(6)$     &           & \cite{Halp14} \\ 
            & $2$   & $1.61(5)$~~  & $0.244(2)$~~~~ &           &                &               &           & \cite{Rodr15} \\
            & $2$   & $1.627(4)$   & $0.229(6)$~~~~ &           &                &               &           & \cite{Kloss12} \\
            & $2$   & $1.61(2)$~~  & $0.2415(15)$   & $0.24(2)$ & $-0.483(3)$    & $1.97(3)$     & $2.00(6)$ & \cite{Odor14,Kelling16b} \\ \hline
{\sc ew}    & $<2$  & $2$          & $(2-d)/4$      & $d/2-1$   & $d/2-1$        & $d$           & $d$       &  \\
            & $2$   & $2$          & $0$(log)$^{\#}$ & $0$      & $0$            & $2$           & $2$       & \cite{Edwards82,Roet06} \\ 
            & $>2$  & $2$          & $0$            & $d/2-1$   & $d/2-1$        & $d$           & $d$       &  \\ \hline
{\rm Arcetri} \hfill $T=T_c$\, 
            & $<2$  & $2$          & $(2-d)/4$      & $d/2-1$   & $d/2-1$        & $3d/2-1$      & $3d/2-1$  & \\ 
            & $2$   & $2$          & $0$(log)$^{\#}$ & $0$      & $0$            & $2$           & $2$       & \cite{Henkel15b} \\ 
            & $>2$  & $2$          & $0$            & $d/2-1$   & $d/2-1$        & $d$           & $d$       & \\[0.19cm]
\hfill $T<T_c$\, & $d$ & $2$       & $1/2$          & $d/2-1$   & $-1$~~         & $d/2-1$       & $d/2-1$   & \\ \hline 
{\sc dle}   & $<1$  & $1$          & $(1-d)/2$      & $d-1$     & $d-1$          & $d$           & $d$       & \\
            & $1$   & $1$          & $0$(log)$^{\#}$ & $0$      & $0$            & $1$           & $1$       & \cite{Krug94,Henkel16a} \\ 
            & $>1$  & $1$          & $0$            & $d-1$     & $d-1$          & $d$           & $d$       & \\ \hline\hline
\end{tabular} \end{center}
~\\
{\footnotesize $^{\#}$For $d=d^*$, one has logarithmic scaling $w(t;L)^2\sim \ln t\: f_w\left( \ln L/\ln t\right)$.} 
\end{sidewaystable}

For comparison, we also list in table~\ref{tab1} values of the non-equilibrium exponents for several other universality classes of interface growth. 
In particular, one sees that for the Edwards-Wilkinson ({\sc ew}) and Arcetri classes, the upper critical dimension $d^*=2$, 
while it is still unknown if a finite value of
$d^*$ exists for the Kardar-Parisi-Zhang ({\sc kpz}) class, see \cite{Barabasi95,Rodr15,Rodrigues15b}. 
Clearly, the stationary exponents $a,b,z$ are the same in the {\sc ew} and Arcetri classes, but the non-equilibrium relaxation
exponents $\lambda_C,\lambda_R$ are different for dimensions $d<d^*$. 
This illustrates the independence of $\lambda_C,\lambda_R$ from those stationary exponents, in agreement with studies
in the non-equilibrium critical dynamics of relaxing magnetic systems. 
On the other hand, for the {\sc kpz} class, a perturbative renormalisation-group analysis shows that $\lambda_C=d$ for $d<2$ \cite{Krech97}. 
For $d>2$, a new strong-coupling fixed point arises and the relaxational properties are still unknown.  
Even for $d=2$, the results of different numerical studies in the {\sc kpz} class are not yet fully consistent. Below, we shall mention recent simulations
with improved extraction of the exponents through precise knowledge of the scaling function from a dynamical symmetry \cite{Kelling16b}. 

We are looking for the dynamical symmetries of the {\sc dle} process. Our main results are as follows.

\noindent
\textcolor{blue}{\bf Theorem.}
{\it The dynamical symmetry of the {\sc dle} process, in $d=1$ space dimension and with $j=0$, is a meta-conformal algebra, in a sense made more precise below and 
isomorphic to the direct sum of three Virasoro algebras without central charge (or `loop-Virasoro algebra').\footnote{Then the maximal finite-dimensional 
Lie sub-algebra isomorphic to $\mathfrak{sl}(2,\mathbb{R})\oplus\mathfrak{sl}(2,\mathbb{R})\oplus\mathfrak{sl}(2,\mathbb{R})$.} 
The Lie algebra generators will be given below in eq.~(\ref{metaconf_inf}), they are non-local in space.  
The general form of the co-variant two-time response function is (with $t>s$)}
\BEA
\lefteqn{R(t,s;r)  =  F_A\:(t-s)^{1-2x}\, \frac{\nu (t-s)}{\nu^2 (t-s)^2 + r^2}} 
\nonumber \\
& & +  F_B\: (t-s)^{1+\psi-2x} \left( \nu^2 (t-s)^2 + r^2\right)^{-(\psi+1)/2} 
\cos\left( (\psi+1)\arctan\left(\frac{r}{\nu (t-s)}\right)-\frac{\pi\psi}{2}\right)~~~~
\label{R_Thm}
\EEA
{\it where $x,\psi$ are real parameters and $F_{A,B}$ are normalisation constants.} \\

\noindent {\bf Comment 1:}
The exact solution (\ref{7R}) of the {\sc dle}-response in $(1+1)D$ is reproduced by (\ref{R_Thm}) 
if one takes $x=\demi$, $\nu>0$, $F_A={\cal C}_0$ and $F_B=0$. 
This illustrates the importance of non-local generators in a specific physical application. 

\noindent {\bf Comment 2:}
The symmetries so constructed are only dynamical symmetries of the so-called {\bf `deterministic part'} of eq.~(\ref{1}), 
which is obtained by setting $T=0$. 
We shall see that the co-variant two-time correlator $C(t,s;r)=0$. 
This agrees with the vanishing of the exact {\sc dle}-correlator (\ref{7C}) in the $T\to 0$ limit
(fix $d\ne 1$ and let first $T\to 0$ and only afterwards $d\to 1$).

This paper presents an exploration of the dynamical symmetries of {\sc dle} process for $d=1$ and is organised as follows. 
In section~2, we introduce the distinction of ortho-conformal and meta-conformal invariance 
and illustrate these notions by several examples, see table~\ref{tab2}. 
In section~3, we explain why none of these local symmetries can be considered as a valid candidate of the dynamical symmetry of the {\sc dle} process. 
Section~4 presents some basic properties on the Riesz-Feller fractional derivative which are used 
in section~5 to explicitly construct the {\em non-local} dynamical symmetry of the {\sc dle} process, thereby
generalising and extending earlier results \cite{Henkel16a}. 
Section~6 outlines the formulation of time-space Ward identities for the computation of covariant $n$-point functions and 
in section~7 the two-point correlator and response are found for the dynamical symmetry of the {\sc dle} process. 
The propositions proven in section~5 and~7 make the theorem~1 more precise and constitute its proof.  
The Lie algebra contraction, in the limit $\nu\to\infty$, and its relationship with the conformal galilean algebra is briefly mentioned. 
This is summarised in table~\ref{tab3}.

\section{Local conformal invariance}

Can one explain the form of the two-time scaling functions of the {\sc dle} process in terms of a dynamical symmetry~? 
To answer such a question, one  must first formulate it more precisely. \\

\noindent {\bf Definition 2:}
{\it The {\bf deterministic part} of the Langevin eq.~(\ref{1}) is obtained when formally setting $\wht{B}=0$.} \\
 
Inspired by Niederer's treatment \cite{Niederer72} of the dynamical symmetries of the free diffusion 
equation,\footnote{The resulting Lie algebra, called {\em Schr\"odinger algebra} by physicists, 
was found by Lie (1881) \cite{Lie1881}. The corresponding continuous symmetries, however, were already known 
to Jacobi (1842/43) \cite{Jacobi1843}. For growing interfaces, the Langevin equation of the {\sc ew} class is the noisy diffusion equation.
Hence its deterministic part, the free diffusion equation, is obviously Schr\"odinger-invariant.}  
we seek dynamical symmetries of the deterministic part of the {\sc dle} process, that is, we look for dynamical symmetries of the non-local
equation $(\mu\partial_t - \nabla_r)\vph=0$, where the non-local Riesz-Feller derivative $\nabla_r$ will be defined below, in section~4. 

Since we see from eq.~(\ref{1}), or the explicit correlators and responses (\ref{6}), that the dynamical exponent $z=1$, 
conformal invariance appears as a natural candidate, where one spatial direction is re-labelled as `time'. 
However, one must sharpen the notion of conformal invariance.\footnote{The names are motivated by the greek prefixes 
$o\vro\theta o$: right, standard and $\mu\vep\tau\alpha$: of secondary rank.} 
For notational simplicity, we now restrict to the case of $1+1$ time-space dimensions, 
labelled by a `time coordinate' $t$ and a `space coordinate' $r$. \\

\noindent {\bf Definition 3:}
{\it (a) A set of {\bf meta-conformal transformations} 
$\mathscr{M}$ is a set of maps $(t,r)\mapsto (t',r')=\mathscr{M}(t,r)$, which may depend analytically on several parameters and form a Lie group. 
The corresponding Lie algebra is isomorphic to the conformal algebra such that the maximal 
finite-dimensional Lie sub-algebra\footnote{Our results on the dynamical symmetries of the {\sc dle} process,
see propositions~3 and~4, require us to present here a more flexible definition than given in \cite{Henkel16a,Henkel16b}.} 
is semi-simple and contains at least a Lie algebra isomorphic to 
$\mathfrak{sl}(2,\mathbb{R})\oplus\mathfrak{sl}(2,\mathbb{R})$. A physical system is 
{\em meta-conformally invariant} if its $n$-point functions transform covariantly under meta-conformal transformations.
(b) A set of {\bf ortho-conformal transformations} $\mathscr{O}$ is a set of meta-conformal transformations 
$(t,r)\mapsto (t',r')=\mathscr{O}(t,r)$, such that (i) the maximal finite-dimensional Lie algebra is isomorphic to 
$\mathfrak{sl}(2,\mathbb{R})\oplus\mathfrak{sl}(2,\mathbb{R})$ and that (ii) angles in the coordinate space of the
points $(t,r)$ are kept invariant. A physical system is 
{\em ortho-conformally invariant} if its $n$-point functions transform covariantly under ortho-conformal transformations. } \\

Ortho-conformal transformations are usually simply called `conformal transformations'. 
We now recall simple examples to illustrate these definitions. See table~\ref{tab2} for a summary.

\begin{table}[tb]
\caption[tab2]{Comparison of local ortho-conformal, conformal galilean and meta-1 conformal invariance, in $(1+1)D$. 
The non-vanishing Lie algebra commutators, the defining equation of the generators, the invariant differential operator $\cal S$ 
and the covariant two-point function is indicated, where applicable. 
Physically, the co-variant quasiprimary two-point function 
$\mathscr{C}_{12}=\langle \vph_1(t,r)\vph_2(0,0)\rangle$ is a correlator, with the constraints $x_1=x_2$ and $\gamma_1=\gamma_2$. 
\label{tab2}} 
\begin{center}\begin{tabular}{||c|lll||} \hline\hline
        & \multicolumn{1}{c}{ortho}                & \multicolumn{1}{c}{galilean}             & \multicolumn{1}{c||}{meta-1} \\ \hline
Lie     & $\left[ X_n, X_m\right] = (n-m) X_{n+m}$ & $\left[ X_n, X_m\right] = (n-m) X_{n+m}$ & $\left[ X_n, X_m\right] = (n-m) X_{n+m}$     \\
algebra & $\left[ X_n, Y_m\right] \:=(n-m)Y_{n+m}$ & $\left[ X_n, Y_m\right] \:=(n-m)Y_{n+m}$ & $\left[ X_n, Y_m\right] \:=(n-m)Y_{n+m}$     \\       
        & $\left[ Y_n,Y_m\right]\:\:=(n-m)X_{n+m}$ & $\left[ Y_n, Y_m\right] \:\:= 0$         & $\left[Y_n, Y_m\right]\:\:=\mu(n-m)Y_{n+m}$  \\[0.14truecm] 
                                                                                                                                                \hline
generators & ~~(\ref{ortho})                       & ~~(\ref{Tcga})                           &  ~~(\ref{Tmeta1})                            \\ \hline
${\cal S}$ & $\partial_t^2 + \partial_r^2$         &                                          & $-\mu\partial_t + \partial_r$                \\[0.14truecm] 
                                                                                                                                                \hline
$\mathscr{C}_{12}$ & $t^{-2x_1}\left( 1 + \left(\frac{r}{t}\right)^2\right)^{-x_1}$   
                   & $t^{-2x_1} \exp\left( -2 \left|\frac{\gamma_1 r}{t}\right|\right)$ 
                   & $t^{-2x_1} \left( 1 + \frac{\mu}{\gamma_1}\left|\frac{\gamma_1 r}{t}\right|\right)^{-2\gamma_1/\mu}$ 
                                                   \\[0.16truecm] \hline \hline     
\end{tabular}\end{center}
\end{table}

\noindent {\bf Example 3:}
In $(1+1)D$, ortho-conformal transformations are analytic
or anti-analytic maps, $z\mapsto f(z)$ or $\bar{z}\mapsto \bar{f}(\bar{z})$, of the complex variables $z =t+\II r$, $\bar{z}=t-\II r$.
The Lie algebra generators are $\ell_n = -z^{n+1}\partial_z$ and $\bar{\ell}_n = -\bar{z}^{n+1}\partial_{\bar{z}}$ with
$n\in\mathbb{Z}$. The conformal Lie algebra is a pair of commuting Virasoro algebras with vanishing central charge \cite{Cartan1909}, viz.
$\left[ \ell_n, \ell_m\right] = (n-m)\ell_{n+m}$. 
In an ortho-conformally invariant physical system, the $\ell_n, \bar{\ell}_n$ act on physical `quasi-primary' \cite{Belavin84}
scaling operators $\phi=\phi(z,\bar{z})=\vph(t,r)$ and contain terms describing  
how these quasi-primary operators should transform, namely
\BEQ \label{ortho}
\ell_n = - z^{n+1}\partial_z - \Delta (n+1) z^n \;\; , \;\; \bar{\ell}_n = -\bar{z}^{n+1}\partial_{\bar{z}} -\overline{\Delta} (n+1) \bar{z}^n
\EEQ
where $\Delta,\overline{\Delta}\in\mathbb{R}$ are the conformal weights of the scaling operator $\phi$. 
The scaling dimension is $x:=x_{\phi}=\Delta+\overline{\Delta}$. Laplace's equation
${\cal S}\phi=4\partial_z \partial_{\bar{z}}\phi=\left(\partial_t^2+\partial_r^2\right)\vph=0$ is a simple example of an ortho-conformally 
invariant system, because of the commutator
\BEQ
\left[ {\cal S}, \ell_n \right]\phi(z,\bar{z}) = -(n+1) z^n {\cal S} \phi(z,\bar{z}) - 4\Delta n(n+1) z^{n-1} \partial_{\bar{z}}\phi(z,\bar{z}) .
\EEQ
This shows that for a scaling operator $\phi$ with $\Delta=\overline{\Delta}=0$, 
the space of solutions of the Laplace equation ${\cal S}\phi=0$ is conformally invariant, 
since any solution $\phi$ is mapped onto another solution $\ell_n\phi$ (or $\bar{\ell}_n\phi$) in the transformed
coordinates. The maximal finite-dimensional sub-group is given by the projective conformal transformations 
$z\mapsto \frac{\alpha z+\beta}{\gamma z+\delta}$ with $\alpha\delta-\beta\gamma=1$; 
its Lie algebra is $\mathfrak{sl}(2,\mathbb{R})\oplus\mathfrak{sl}(2,\mathbb{R})$. Two-point functions of quasi-primary scaling operators read 
\BEQ
\mathscr{C}_{12}(t_1,t_2;r_1,r_2) := \langle \phi_1(z_1,\bar{z}_1) \phi_2(z_2,\bar{z}_2)\rangle = \langle \vph_1(t_1,r_1) \vph_2(t_2,r_2)\rangle.
\EEQ 
Their ortho-conformal covariance implies the projective Ward identities $X_n \mathscr{C}_{12}=Y_n \mathscr{C}_{12}=0$ for
$n=\pm 1,0$ \cite{Belavin84}. For scalars, such that $\Delta_i=\overline{\Delta}_i=x_i$, this gives, up to the
normalisation ${\cal C}_0$ \cite{Polyakov70}
\BEQ \label{Cortho}
\mathscr{C}_{12}(t_1,t_2;r_1,r_2) = {\cal C}_0\, \delta_{x_1,x_2} \left( (t_1-t_2)^2 + (r_1-r_2)^2 \right)^{-x_1}.
\EEQ
Below, we often use the basis $X_n := \ell_n +\bar{\ell}_n$ and $Y_n := \ell_n -\bar{\ell}_n$, see
also table~\ref{tab2}. \\

\noindent {\bf Example 4:}
An example of meta-conformal transformations in $(1+1)D$ reads \cite{Henkel02}
\BEA
X_n   &=& -t^{n+1}\partial_t-\mu^{-1}[(t+\mu r)^{n+1}-t^{n+1}]\partial_r-(n+1)xt^n- (n+1)\frac{\gamma}{\mu}[(t+\mu r)^{n}-t^{n}]\nonumber\\
Y_{n} &=& -(t+\mu r)^{n+1}\partial_r- (n+1)\gamma (t+\mu r)^{n}
\label{Tmeta1}
\EEA
with $n\in\mathbb{Z}$. Herein, $x,\gamma$ are the scaling dimension and the `rapidity' of the scaling operator 
$\vph=\vph(t,r)$ on which these generators act.  
The constant $1/\mu$ has the dimensions of a velocity. The Lie algebra $\langle X_n, Y_n\rangle_{n\in\mathbb{Z}}$ 
is isomorphic to the conformal Lie algebra \cite{Henkel15c}, see table~\ref{tab2}, where it is called {\bf meta-1 conformal invariance}. 
If $\gamma=\mu x$, the generators (\ref{Tmeta1}) act as dynamical symmetries on the equation ${\cal S}\vph=(-\mu\partial_t + \partial_r)\vph=0$. 
This follows from the only non-vanishing commutators of the Lie algebra with $\cal S$, namely $\left[{\cal S},X_0\right]\vph = -{\cal S}\vph$ and 
$\left[{\cal S},X_{1}\right]\vph = -2t{\cal S}\vph+2(\mu x-\gamma)\vph$. 
The formulation of the meta-1 conformal Ward identities does require some care, 
since already the two-point function turns out to be a non-analytic function of the time- and space-coordinates. 
It can be shown that the covariant two-point correlator is \cite{Henkel16b}
\BEQ \label{Cmeta1}
\mathscr{C}_{12}(t_1,t_2;r_1,r_2) = {\cal C}_0\: \delta_{x_1,x_2} \delta_{\gamma_1,\gamma_2} \left| t_1-t_2\right|^{-2x_1} 
\left( 1 + \frac{\mu}{\gamma_1} \left| \gamma_1 \frac{r_1-r_2}{t_1-t_2}\right| \right)^{-2\gamma_1/\mu}.
\EEQ  
~ \\

Although both examples have $z=1$ and isomorphic Lie algebras, the explicit two-point functions (\ref{Cortho}) and (\ref{Cmeta1}), as well as the
invariant equations ${\cal S}\vph=0$, are different,\footnote{The representation (\ref{Tmeta1}) can be extended to produce dynamical symmetries of the 
$(1+1)D$ Vlasov equation \cite{Stoimenov15}.} see also table~\ref{tab2}. 
\textcolor{black}{That the form of two-point functions depends mainly on the representation and not so much on the Lie algebra, 
is not a phenomenon restricted to the conformal algebra. Similarly, for the so-called {\em Schr\"odinger algebra} 
at least threee distinct representations with different forms of the two-point function are known \cite{Henkel06b}.}
\\

\noindent {\bf Example 5:}
Taking the limit $\mu\to 0$ in the meta-conformal representation (\ref{Tmeta1}) produces the generators
\BEA
X_n   &=& -t^{n+1}\partial_t-(n+1) t^n r\partial_r-(n+1)xt^n- (n+1)n \gamma t^{n-1} r \nonumber\\
Y_{n} &=& -t^{n+1}\partial_r- (n+1)\gamma t^{n}
\label{Tcga}
\EEA
of the {\bf conformal Galilean algebra} ({\sc cga}) in $(1+1)D$ 
\cite{Havas78,Henkel97,Negro97,Negro97b,Barnich07,Barnich07b,Bagchi09b,Duval09,Cherniha10,Hosseiny10b,Zhang10}.
Its Lie algebra is obtained by standard contraction of the conformal Lie algebra, 
see table~\ref{tab2}.\footnote{\textcolor{black}{It can be shown \cite{Cherniha10} that there are no $\mbox{\sc cga}$-invariant scalar equations 
(in the classical Lie sense). However, if one considers the
Newton-Hooke extension of the $\mbox{\sc cga}$ on a curved de Sitter/anti-de Sitter space (whose flat-space limit is not isomorphic to the $\mbox{\sc cga}$), 
non-linear representations have been used  to find non-linear invariant equations, related to the Pais-Uhlenbeck oscillator,  
see \cite{CherMastKriv16} and refs. therein.}} 
Hence the {\sc cga} is {\em not}
a meta-conformal algebra, although $z=1$. The co-variant two-point correlator
can either be obtained from the generators (\ref{Tcga}), using techniques similar to those applied in the above example of meta-conformal
invariance \cite{Henkel15,Henkel15c}, or else by letting $\mu\to 0$ in (\ref{Cmeta1}). Both approaches give  
\BEQ \label{Ccga}
\mathscr{C}_{12}(t_1,t_2;r_1,r_2) = {\cal C}_0\: \delta_{x_1,x_2} \delta_{\gamma_1,\gamma_2} \left| t_1-t_2\right|^{-2x_1} 
\exp \left( -2 \left| \gamma_1 \frac{r_1-r_2}{t_1-t_2}\right| \right)
\EEQ
Clearly, this form is different from both ortho- and meta-1 conformal invariance.  \\

The non-analyticity of the correlators (\ref{Cmeta1}), and especially (\ref{Ccga}), in general overlooked in the literature, 
is {\em required} in order to achieve $\mathscr{C}_{12}\to 0$ for large time- or space-separations, 
viz. $t_1-t_2\to\pm\infty$ or $r_1-r_2\to \pm\infty$. 

All two-point functions (\ref{Cortho},\ref{Cmeta1},\ref{Ccga}) have indeed the symmetries 
$\mathscr{C}_{12}(t_1,t_2;r,r)=\mathscr{C}_{21}(t_2,t_1;r,r)$ and 
$\mathscr{C}_{12}(t,t;r_1,r_2)=\mathscr{C}_{21}(t,t;r_2,r_1)$, under permutation $\vph_1 \leftrightarrow \vph_2$ of the two scaling operators, 
as physically required for a correlator. \textcolor{black}{The shape of the scaling functions of these three 
two-point functions is compared in figure~\ref{fig3}. 
\begin{figure}[tb]
\centerline{\includegraphics[scale=0.6]{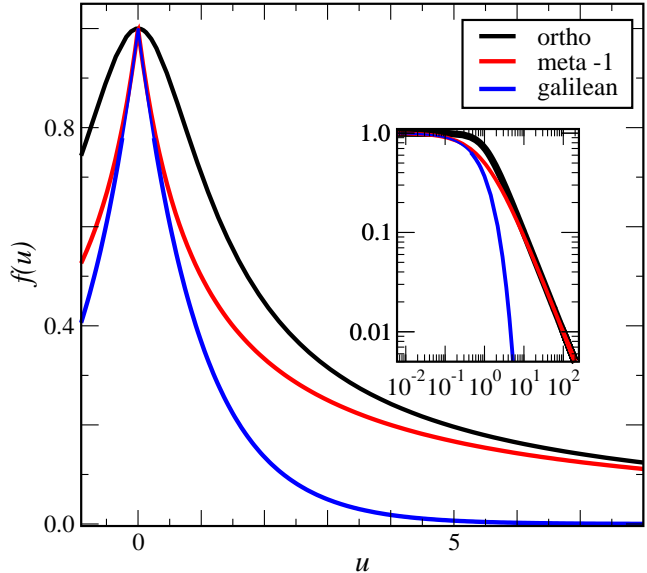}} 
\caption[fig3]{Scaling function $f(u)$ of the covariant two-point correlator $\mathscr{C}(t,r)=t^{-2x_1}f(r/t)$, 
over against the scaling variable $u=r/t$, for ortho-, meta-1- and Galilean-conformal invariance, 
eqs.~(\ref{Cortho},\ref{Cmeta1},\ref{Ccga}) respectively, where $x_1=\gamma_1=\demi$ and $\mu=1$.  
\label{fig3}}
\end{figure}
In particular, the non-analyticity of the meta- and Galilean conformal invariance at $u=0$ is clearly seen, in contrast to ortho-conformal invariance, while
for $u\to\infty$, the slow algebraic decay of ortho- and meta-conformal invariance is distinct from the exponential decay of conformal Galilean invariance.
This illustrates the variety of possible forms already for $z=1$. Below, we shall find another form of (meta-)conformal invariance, 
different from all forms displayed in figure~\ref{fig3}.} 

\section{Impossibility of a local meta-conformal invariance of the {\sc dle} process}

Can one consider these several variants of conformal invariance, which have $z=1$ and are realised in terms 
of {\it local} first-order differential operators, as a valid dynamical symmetry of the {\sc dle} process in $1+1$ dimensions~? 
The answer turns out to be negative:  
\begin{enumerate}
\item The deterministic part of the {\sc dle} Langevin equation (\ref{1}) is distinct from the simple invariant equations ${\cal S}\vph=0$ 
of either ortho- or meta-1 conformal invariance. 

For analogy, consider briefly {\it Schr\"odinger-invariant systems} with a Langevin equation of the form ${\cal S}\vph=\left(2\nu T\right)^{1/2}\eta$, 
where $\eta=\frac{\D B}{\D t}$ is a white noise of unit variance, 
and such that
the Schr\"odinger algebra is a dynamical symmetry of the noise-less equation (deterministic part) ${\cal S}\vph_0=0$. 
Then, the Bargman super-selection rules \cite{Bargman54} which follow from the combination of spatial 
translation-invariance and Galilei-invariance with $z=2$, imply exact relations between averages of the
full noisy theory and the averages calculated from its deterministic part \cite{Picone04}. In particular, the two-time response
function of the full noisy equation $R(t,s;\vec{r})=R_0(t,s;\vec{r})$, is identical to 
the response $R_0$ found when the noise is turned off and computed from the dynamical Schr\"odinger symmetry \cite{Picone04,Henkel10}.  

We shall assume here that an analogous result can be derived also for the {\sc dle} Langevin equation, although this has not yet been done. 
It seems plausible that such a result should exist, since in the example (\ref{6R},\ref{7R}) of the {\sc dle} process, 
the two-time response $R$ is independent of $T$ (which characterises the white noise), as it is the case for Schr\"odinger-invariance. 
\item The explicit response function (\ref{7R}) of the {\sc dle} process is distinct from the predictions (\ref{Cortho},\ref{Cmeta1},\ref{Ccga}), 
see also table~\ref{tab2}. 
The form of the meta-1 conformal two-point function (\ref{Cmeta1}), 
is clearly different for finite values of the scaling variable $v=(r_1-r_2)/(t_1-t_2)$, and similarly for the galilean conformal case (\ref{Ccga}). 
The ortho-conformal two-point function (\ref{Cortho})
looks to be much closer, with the choice $x_1=\demi$ and the scale factor fixed to $\nu=1$, were it not for the extra factor $\nu(t-s)$. 

On the other hand, the two-time {\sc dle} correlator (\ref{7C}) does not agree with (\ref{Cortho}) either, but might be similar to a two-point function
computed in a semi-infinite space $t\geq 0$, $r\in\mathbb{R}$ with a boundary at $t=0$. 
\end{enumerate}
Looking for dynamical symmetries of the deterministic part of the {\sc dle} Langevin equation (\ref{1}), in $1+1$ dimensions, 
the first test will be the computation of the two-time response function 
$R(t,s;\vec{r})$. By contrast, we shall show that the two-time correlator $C$ cannot be found in this way. 
Indeed, its `deterministic' contribution vanishes: $C_0(t,s;\vec{r})=0$.

\section{Riesz-Feller fractional derivative}

Formulating (\ref{1}) in direct space requires the Riesz-Feller fractional derivative \cite{Samk93,Nezza12,Cinti15} of order $\alpha$. \\

\noindent {\bf Definition 4:} 
{\it For functions $f(r)$ of a single variable $r\in\mathbb{R}$, the {\bf Riesz-Feller derivative, of order $\alpha$}, is}  
\BEQ \label{RFderivee}
\nabla_r^{\alpha} f(r) := \frac{\II^{\alpha}}{\sqrt{2\pi\,}\,} \int_{\mathbb{R}} \!\D k\: |k|^{\alpha}\: e^{\II kr}\, \wht{f}(k) 
= \frac{\II^{\alpha}}{2\pi} \int_{\mathbb{R}^2} \!\D k\D x\: |k|^{\alpha}\: e^{\II k(r-x)}\, f(x)
\EEQ
{\it where $\wht{f}(k)$ denotes the Fourier transform of $f(r)$. For brevity, we often write 
$\nabla_{{r}} = \nabla_{r}^1$ and distinguish it from the standard derivative $\partial_{{r}}$.} \\ 

\noindent
{\bf Lemma 1:} \cite[Prop. 3.6]{Nezza12} 
{\it Let $f\in H^{\alpha/2}(\mathbb{R}) = \left\{ f\in L^2(\mathbb{R})\left| 
\int_{\mathbb{R}} \D k\, |\wht{f}(k)|^2 (1+|k|^{2})^{\alpha/2}<\infty\right.\right\}$, a fractional Sobolev space. 
For $0<\alpha<2$, the Riesz-Feller derivative $\nabla_r^{\alpha}f(r)$ exists.} \\ 

\noindent
{\bf Lemma 2:} \cite[app. J.2]{Henkel10},\cite{Henkel16a} 
{\it The following formal properties hold true,  where $\alpha, \beta, q,\mu$ are constants}
\BEA
\nabla_r^{\alpha}\nabla_r^{\beta} f(r) = \nabla_r^{\alpha+\beta} f(r) \;\; &,& \;\;
\left[ \nabla_r^{\alpha}, r \right] f(r) = \alpha \partial_r \nabla_r^{\alpha-2} f(r) \;\; , \;\; 
\nabla_r^{\alpha} f(\mu r) = |\mu|^{\alpha} \nabla_{\mu r}^{\alpha} f(\mu r) \nonumber \\
\nabla_r^{\alpha} e^{\II q r} = \left(\II |q|\right)^{\alpha} e^{\II q r} \;\; &,& \;\; 
\left(\wht{\nabla_r^{\alpha} f(r)}\right)(q) = \left(\II |q|\right)^{\alpha} \wht{f}(q) \;\; , \;\;
\nabla_r^2 f(r) = \partial_r^2 f(r)
\label{16}
\EEA

\noindent Lemma 2 follows directly from the definition (\ref{RFderivee}). 
The analogy with the rules of the ordinary derivative $\partial_{r}^n$, with $n\in\mathbb{N}$, applied to exponentials $e^{\II q r}$ and to
Fourier transforms, motivated our choice of (complex) normalisation in (\ref{RFderivee}). Later, we shall also need the object
$\partial_r \nabla_r^{-1}$, which is formally written as 
\BEQ \label{16bis}
\partial_r \nabla_r^{-1} f(r) = \frac{1}{\sqrt{2\pi\,}\,} \int_{\mathbb{R}} \!\D k\: e^{\II k r}\, \sign (k)\,  \wht{f}(k) 
\EEQ  
but is best considered via its Fourier transform, viz.  $\left(\wht{\partial_r \nabla_r^{-1} f(r)}\right)(q) = \sign(q) \wht{f}(q)$. 
This is well-defined, since $f\in H^{\alpha/2}(\mathbb{R})\subset L^2(\mathbb{R})$ and because of Plancherel's theorem. 
The Fourier transform of $\sign(q)$ is a distribution \cite[eq. (2.3.17)]{Gelfand}. \\ 

\noindent {\bf Corollary 1:} \label{co1}
One has the following formal commutator identities 
\BEA 
&& \left[ \nabla_r, r^n\right] = nr^{n-1}\partial_r\nabla_r^{-1} \;\;,\;\; 
\left[ r^2\nabla_r,r\partial_r\right] = -r^2\nabla_r \;\;,\;\; 
\left[ r\partial_r\nabla_r^{-1}, r \nabla_r \right] = - r \;\;,
\nonumber \\
&& \left[ \nabla_r, \partial_r \right]= \left[ r\partial_r , r\nabla_r \right] = \left[r, \partial_r \nabla_r^{-1} \right] 
= \left[ r^n \partial_r, \partial_r \nabla_r^{-1} \right]=\left[ r^n\nabla_r , \partial_r\nabla_r^{-1}\right]= 0 \;\;.
\label{18}
\EEA

\noindent
{\bf Proof:} 
Most of these identities are immediate consequences of (\ref{RFderivee},\ref{16}). The first one is proven by induction, for all $n\in\mathbb{N}$. 
We only detail here the computation of the third one. Formally, one has  
$\left[ r\partial_r\nabla_r^{-1}, r \nabla_r \right] = -r \left( \partial_r \nabla_r^{-1}\right)^2$. 
Using (\ref{16bis}), and its Fourier transform, gives
\BD
\left( \partial_r \nabla_r^{-1}\right)^2 f(r) = \int_{\mathbb{R}} \frac{\D k}{\sqrt{2\pi\,}\,} \: e^{\II k r} 
\sign(k) \left( \wht{\partial_r \nabla_r^{-1}f(r)}\right)(k) 
= \int_{\mathbb{R}} \frac{\D k}{\sqrt{2\pi\,}\,}\: e^{\II k r}  \left(\sign(k)\right)^2 \wht{f}(k) = f(r)
\ED
which establishes the assertion. \hfill q.e.d \\

\noindent {\bf Lemma 3:} \cite{Nezza12,Sethuraman16} 
{\it If $0<\alpha<2$, one has for either $f\in H^{\alpha/2}(\mathbb{R})$ or else $f\in\mathscr{S}(\mathbb{R})$, the Schwartz space of smooth, 
rapidly decreasing functions, for almost all $r\in\mathbb{R}$}
\BEQ \label{RFCauchy}
\nabla_r^{\alpha} f(r) = \frac{1-e^{\II\pi\alpha}}{4\pi \II}\Gamma(\alpha+1) 
\int_{\mathbb{R}} \!\frac{\D y}{|y|^{\alpha+1}} \left[ f(r+y) - 2f(r) + f(r-y) \right]
\EEQ

\noindent
{\bf Proof:} 
One may check the identity with the definition (\ref{RFderivee}) in Fourier space. 
Indeed, writing for the moment $s_{\alpha}$ for the constant prefactor before the integral in (\ref{RFCauchy})
\BEA
\left( \wht{\nabla_r^{\alpha} f(r)} \right)(q) &=& \frac{1}{\sqrt{2\pi\,}} \int_{\mathbb{R}} \!\D r\: e^{-\II q r}\: \nabla_r^{\alpha} f(r) 
\nonumber \\
&=& \frac{s_{\alpha}}{2\sqrt{2\pi\,}} \int_{\mathbb{R}} \frac{\D y}{|y|^{\alpha+1}} 
\int_{\mathbb{R}} \!\D r\: \left[ e^{-\II q r} f(r+y) - 2 e^{-\II q r} f(r) + e^{-\II q r} f(r-y)\right]
\nonumber \\
&=& \frac{s_{\alpha}}{2} \int_{\mathbb{R}} \frac{\D y}{|y|^{\alpha+1}} \wht{f}(q) \left[ e^{\II q y} - 2 + e^{-\II q y} \right]
\nonumber \\
&=& 2s_{\alpha} \Gamma(-\alpha) \cos \frac{\pi \alpha}{2} \: |q|^{\alpha} \wht{f}(q) \nonumber
\EEA
and consistency with the 5$^{\rm th}$ relation in (\ref{16}) leads to 
$2 s_{\alpha} \Gamma(-\alpha)\cos \frac{\pi \alpha}{2} \stackrel{!}{=} \II^{\alpha} = e^{\II\pi \alpha/2}$ 
which reproduces the constant in (\ref{RFCauchy}). 
The well-known theorem on the identity of functions in Fourier space gives the assertion. \hfill ~q.e.d.\\

\noindent If $f\in\mathscr{S}(\mathbb{R})$, $\nabla_r^{\alpha}f(r)$ can be defined beyond the interval $0<\alpha<2$.

\section{Non-local meta-conformal generators}

The deterministic part of (\ref{1}) becomes ${\cal S}\vph = \left(-\mu\partial_r + \nabla_r\right)\vph=0$, where $\mu^{-1}=\II \nu$. 
Following earlier studies \cite{Henkel02}, it seems physically reasonable that the Lie algebra of dynamical symmetries should at least contain
the generators of time translations $X_{-1}=-\partial_t$, dilatations $X_0 = -t\partial_t - \frac{r}{z}\partial_r -\frac{x}{z}$ 
and space translations $Y_{-1}=-\partial_r$. 
It turns out, however, if one wishes to construct a generator of generalised Galilei transformations as a dynamical symmetry, 
the non-local generator $-\nabla_r$ automatically arises, see below and \cite[ch. 5.3]{Henkel10}. 
It is still an open problem how to close these generators into a Lie algebra, for $z\ne 2$
and beyond the examples listed above in section~2. 

This difficulty motivates us to start with the choice of a {\it non-local} spatial translation operator $Y_{-1}=-\mu^{-1}\nabla_r$. 
Here indeed, a closed Lie algebra can be found. \\

\noindent
{\bf Proposition 1:} \cite{Henkel16a} {\it Define the following generators} 
\BEA
X_{-1} &=& \!\!-\partial_t \;\; , \;\; X_0 = -t \partial_t - r \partial_r -x 
\;\; , \;\;
X_1 = -t^2\partial_t -2tr\partial_r -\mu r^2\nabla_r -2xt -2\gamma r\partial_r \nabla_r^{-1}
\label{1Part} 
\\
Y_{-1} &=& \!\!-\frac{1}{\mu}\nabla_r \; , \;\; Y_0 = -\frac{1}{\mu}t\nabla_r - r \partial_r -\frac{\gamma}{\mu} 
\; , \; 
Y_1 = -\frac{1}{\mu}t^2\nabla_r -2 tr \partial_r -\mu r^2 \nabla_r -2\frac{\gamma}{\mu} t -2\gamma r \partial_r \nabla_r^{-1}
\nonumber 
\EEA
{\it where the constants $x=x_{\vph}$ and $\gamma=\gamma_{\vph}$, respectively, are the scaling dimension and rapidity of
the scaling operator $\vph=\vph(t,r)$ on which these generators act. The six generators (\ref{1Part}) 
obey the commutation relations of a meta-conformal Lie algebra, isomorphic to 
$\mathfrak{sl}(2,\mathbb{R})\oplus\mathfrak{sl}(2,\mathbb{R})$} 
\BEQ \label{metaconf}
\left[ X_n, X_m\right] =(n-m) X_{n+m} \;\; , \;\; \left[ X_n, Y_m\right] =(n-m) Y_{n+m} \;\; , \;\; \left[ Y_n, Y_m\right] = (n-m) Y_{n+m}
\EEQ

\noindent {\bf Proposition 2:} \cite{Henkel16a}
{\it The generators (\ref{1Part}) obey the commutators} 
\BEQ
\left[ {\cal S}, Y_n \right]\vph = \left[ {\cal S}, X_{-1} \right]\vph = 0 \;\; , \;\;
\left[ {\cal S}, X_0 \right]\vph = - {\cal S}\vph \;\; , \;\;
\left[ {\cal S}, X_1 \right]\vph = -2t{\cal S}\vph +2(\mu x-\gamma)\vph
\EEQ
{\it with the differential operator ${\cal S}=-\mu\partial_t +\nabla_r$ and thus form a Lie algebra of meta-conformal dynamical symmetries 
(of the deterministic part ${\cal S}\vph=0$) of the {\sc dle} Langevin equation (\ref{1}), if only $\gamma=x\mu$.} \\

\noindent Propositions 1 and 2 are established by direct formal calculations. 
The non-local generators $X_{1}, Y_{0,1}$ in (\ref{1Part}) do not generate simple local changes of the coordinates $(t,r)$, 
in contrast to all examples of section~2. 
Finding a clear geometrical interpretation of the generators (\ref{1Part}) remains an open problem. 

This meta-conformal symmetry algebra can be considerably enlarged. \\

\noindent
{\bf Proposition 3:}
{\it Consider the generators (\ref{1Part}) and furthermore define}
\BEQ \label{1PartZ}
Z_{-1} = -\frac{1}{\mu}\partial_r \; , \;\; Z_0 = -\frac{1}{\mu}t\partial_r - r \nabla_r -\frac{\gamma}{\mu} \partial_r\nabla_r^{-1} 
\; , \; 
Z_1 = -\frac{1}{\mu}t^2\partial_r -2 tr \nabla_r -\mu r^2 \partial_r -2\frac{\gamma}{\mu} t \partial_r\nabla_r^{-1} -2\gamma  r ~~~
\EEQ
{\it These generators are dynamical symmetries of the {\sc dle} process, since $\left[{\cal S},Z_n\right]=0$ and they extend the meta-conformal
Lie algebra (\ref{metaconf}) as follows} 
\BEQ \label{metaconf_bis}
\left[ X_n, Z_m\right] =(n-m) Z_{n+m} \;\; , \;\; \left[ Y_n, Z_m\right] =(n-m) Z_{n+m} \;\; , \;\; \left[ Z_n, Z_m\right] = (n-m) Y_{n+m}
\EEQ

\noindent Although $Z_{-1}$ generates local spatial translations, the transformations obtained from $Z_{0,1}$ are non-local. 
In what follows, we write $\xi:=\gamma/\mu$ for the second, independent scaling dimension of $\vph$. 

\noindent
{\bf Corollary 2:}
Define the generators $B_n^{\pm} = \demi \left( Y_n \pm Z_n\right)$, $n\in\{-1,0,1\}$. Then the non-vanishing commutators of the Lie algebra 
(\ref{metaconf},\ref{metaconf_bis}) take the form 
\BEQ \label{metaconf_ter}
\left[ X_n, X_m\right] =(n-m) X_{n+m} \;\; , \;\; \left[ X_n, B_m^{\pm}\right] =(n-m) B_{n+m}^{\pm} \;\; , \;\; 
\left[ B_n^{\pm}, B_m^{\pm}\right] = (n-m) B_{n+m}^{\pm}
\EEQ
The $B_n^{\pm}$ are dynamical symmetries of the {\sc dle} process, since $\left[{\cal S},B_n^{\pm}\right]=0$. \\

\noindent
{\bf Corollary 3:} \label{co3}
Define the generators $A_n = X_n -\left( B_n^+ + B_n^-\right)=X_n - Y_n$, $n\in\{-1,0,1\}$. Then the non-vanishing commutators of
the Lie algebra (\ref{metaconf_ter}) are
\BEQ \label{metaconf_qua}
\left[ A_n, A_m\right] =(n-m) A_{n+m} \;\; , \;\; \left[ B_n^{\pm}, B_m^{\pm}\right] = (n-m) B_{n+m}^{\pm}
\EEQ
This Lie algebra of dynamical symmetries of the deterministic part of the {\sc dle} equation (\ref{1}) is isomorphic to the 
direct sum $\mathfrak{sl}(2,\mathbb{R})\oplus\mathfrak{sl}(2,\mathbb{R})\oplus\mathfrak{sl}(2,\mathbb{R})$. \\

In this last choice of basis, all generators contain non-local terms. Their form, in corollary~3, is suggestive for the explicit construction
of an infinite-dimensional extension of the above Lie algebra. \\

\noindent
{\bf Proposition 4:}
{\it Construct the generators, for all $n\in\mathbb{Z}$ and $x,\xi$ constants} 
\BEA  \label{metaconf_inf}
A_n &=& -t^{n+1} \left( \partial_t - \nabla_r \right) - (n+1) \left( x - \xi\right) t^n \nonumber \\
B_n^{\pm} &=& -\demi\left( t\pm r\right)^{n+1} \left( \nabla_r \pm \partial_r\right) - \frac{n+1}{2}\xi \left( t\pm r\right)^{n}
\left( 1 \pm \partial_r \nabla_r^{-1} \right)
\EEA
{\it Their non-vanishing commutators are given by (\ref{metaconf_qua}), for $n,m\in\mathbb{Z}$. Their Lie algebra is isomorphic to the direct sum
of three Virasoro algebras with vanishing central charges. They are also dynamic symmetries of the deterministic equation
${\cal S}\vph = \left(-\partial_t + \nabla_r\right)\vph=0$ of the {\sc dle} process, provided that $x=\xi$, because of the commutators}
\BEQ
\left[ {\cal S}, A_n \right] = -(n+1) t^n {\cal S} + (n+1)n\left( x-\xi\right) t^{n-1} \;\; , \;\;
\left[ {\cal S}, B_n^{\pm} \right] = 0 
\EEQ

\noindent
{\bf Proof:} For $n=\pm 1,0$, the generators (\ref{metaconf_inf}) are those given above in (\ref{1Part},\ref{1PartZ}), 
using $\xi=\gamma/\mu$ and rescaling $\mu\mapsto 1$. 
One generalises the first identity (\ref{18}) in the corollary~\ref{co1} to the following form, with $n\in\mathbb{N}$
\BD
\left[ \nabla_r , \left( \alpha \pm r \right)^n \right] = \pm n \left( \alpha \pm r \right)^{n-1} \partial_r \nabla_r^{-1}
\ED 
where $\alpha$ is a constant. The assertions now follow by straightforward formal calculations, using (\ref{16},\ref{18}). \hfill q.e.d. \\

This is the {\sc dle}-analogue of the ortho- and meta-1 conformal invariances, 
respectively, of the Laplace equation and of simple ballistic transport, 
as treated in examples~3 and~4. In table~\ref{tab3}, it is called `meta-2 conformal'. 
It clearly appears that both local and  non-local spatial translations are needed for realising the full
dynamical symmetry of the {\sc dle} process, which we call {\bf erosion-Virasoro algebra} and denote by $\mathfrak{ev}$. 
The infinite-dimensional Lie algebra $\mathfrak{ev}$ is built from {\em three} commuting Virasoro algebras, 
and has the maximal finite-dimensional Lie sub-algebra 
$\mathfrak{sl}(2,\mathbb{R})\oplus\mathfrak{sl}(2,\mathbb{R})\oplus\mathfrak{sl}(2,\mathbb{R})$.  
The scaling operators $\vph=\vph(t,r)$ on which these generators act are characterised by two independent 
{\em `scaling dimensions'} $x=x_{\vph}$ and $\xi=\xi_{\vph}$. 
By analogy with conformal galilean invariance \cite{Ovsienko98}, one expects that three independent central charges of the 
Virasoro type should appear if the algebra (\ref{metaconf_qua}) will be quantised. 
Additional physical constraints (e.g. unitarity) may reduce the number of independent central charges. 

\section{Ward identities for co-variant quasi-primary $n$-point functions}

A basic application of dynamic time-space symmetries is the derivation of co-variant $n$-point functions. 
Adapting the corresponding definition from (ortho-)conformal invariance \cite{Belavin84}, 
a scaling operator $\vph=\vph(t,r)$ is called {\bf quasi-primary}, if it transforms co-variantly
under the action of the generators of the maximal finite-dimensional sub-algebra of $\mathfrak{ev}$. 
A {\em primary} scaling operator transforms co-variantly under the
action of all generators of $\mathfrak{ev}$. In this work, we consider examples of $n$-point functions of quasi-primary scaling operators. 

In the physical context of non-equilibrium dynamics, such $n$-point functions can either be correlators, 
such as $\langle \vph(t,r)\vph(t',r')\rangle$, or response functions
$\langle \vph(t,r)\wit{\vph}(t',0)\rangle = \left.\frac{\delta \langle\vph(t,r)\rangle}{\delta j(t',0)}\right|_{j=0}$, 
which can be formally rewritten as a correlator
by using the formalism of Janssen-de Dominicis theory \cite{Taeuber14} which defines the response operator 
$\wit{\vph},$ conjugate to the scaling operator $\vph$. 

Proceeding in analogy with ortho-conformal  and Schr\"odinger-invariance \cite{Polyakov70,Belavin84,Francesco97,Henkel10}, 
the quasi-primary $\mathfrak{ev}$-Ward identities are obtained from the explicit form of 
the Lie algebra generators (\ref{1Part},\ref{1PartZ}), generalised to $n$-body generators. 
In order to do so, we assign a {\bf signature} $\vep=\pm 1$ to each scaling operator. 
We choose the convention that $\vep_i=+1$ for scaling operators $\vph_i$ and $\vep_i=-1$ for
response operators $\wit{\vph}_i$. In order to prepare a later application to the conformal galilean algebra, 
to be obtained from a Lie algebra contraction, we also multiply the generators
$Y_i, Z_i$ by the scale factor $\mu$. The $n$-body generators then read
\BEA
X_{-1} \:=\: X_{-1}^{[n]} &=& \sum_i \left[ -\partial_i\right] \hspace{1.0truecm},\hspace{1.0truecm} 
X_0  \:=\: X_0^{[n]} \:=\: \sum_i \left[ -t_i\partial_i - r_iD_i - x_i \right] \nonumber \\
X_1 \:=\: X_1^{[n]} &=& \sum_i \left[ -t_i^2\partial_i -2t_ir_i D_i -\mu \vep_i r_i^2 \nabla_i -2x_i t_i -2\mu\xi_i \vep_i r_i D_i \nabla_i^{-1}\right]
\nonumber \\
Y_{-1} \:=\: Y_{-1}^{[n]} &=& \sum_i \left[ -\vep_i \nabla_i \right] \hspace{0.6truecm},\hspace{1.0truecm} 
Y_0 \:=\: Y_0^{[n]} \:=\: \sum_i \left[ -\vep_i t_i \nabla_i -\mu r_i D_i -\mu\xi_i \right] \label{2Part} \\
Y_1 \:=\: Y_1^{[n]} &=& \sum_i \left[ -\vep_i \left( t_i^2 +\mu^2 r_i^2\right)\nabla_i -2\mu t_i r_i D_i  
                                      -2\mu\xi_i t_i -2\mu^2\xi_i \vep_i r_i D_i \nabla_i^{-1}\right]
\nonumber \\
Z_{-1} \:=\: Z_{-1}^{[n]} &=& \sum_i \left[ - D_i \right] \hspace{0.9truecm},\hspace{1.0truecm} 
Z_0 \:=\: Z_0^{[n]} \:=\: \sum_i \left[ -t_i D_i - \vep_i r_i \nabla_i -\mu\xi_i D_i \nabla_i^{-1} \right] \nonumber \\
Z_{1} \:=\: Z_1^{[n]} &=& \sum_i \left[ -\left( t_i^2 +\mu^2 r_i^2\right) D_i -2\vep_i \mu t_i r_i \nabla_i 
                                        -2\mu\xi_i r_i -2\mu\xi_i \vep_i t_i D_i \nabla_i^{-1} \right]
\nonumber 
\EEA
with the short-hands $\partial_i=\frac{\partial}{\partial t_i}$, $D_i = \frac{\partial}{\partial r_i}$ and $\nabla_i = \nabla_{r_i}$. 
It can be checked that the generators (\ref{2Part}) obey the meta-conformal Lie algebra of the {\sc dle} process. Define the $(n+m)$-point function
\BEA 
\mathscr{C}_{n,m}&=&\mathscr{C}_{n,m}(t_1,\ldots,t_{n+m};r_{1},\ldots,r_{n+m}) \nonumber \\
&=& \left\langle \vph_1(t_1,r_1)\cdots\vph_n(t_n,r_n) \wit{\vph}_{n+1}(t_{n+1},r_{n+1})\cdots\wit{\vph}_{n+m}(t_{n+m},r_{n+m})\right\rangle
\label{Cnm}
\EEA
of quasi-primary scaling and response operators. Their co-variance is expressed through the quasi-primary Ward identities, for $k=\pm 1,0$
\BEQ
X_{k}^{[n+m]} \mathscr{C}_{n,m}=Y_{k}^{[n+m]}\mathscr{C}_{n,m}=Z_{k}^{[n+m]}\mathscr{C}_{n,m}=0,
\EEQ 
The solution of this set of (linear) differential equations gives the sought $(n+m)$-point function $\mathscr{C}_{n,m}$. 

\section{Co-variant two-time correlators and responses}

In order to illustrate the procedure outlined in section~6, we shall apply it to the two-point functions. 

\noindent
{\bf Proposition 5:}
{\it Any two-point correlator $\mathscr{C}_{2,0}(t_1,t_2;r_1,r_2)=\langle \vph_1(t_1,r_1)\vph_2(t_2,r_2)\rangle$, built from
$\mathfrak{ev}$-quasi-primary scaling operators $\vph_i$, vanishes.} 

\noindent
{\bf Proof:} Time-translation-invariance, expressed by $X_{-1}\mathscr{C}_{2,0}=0$, 
implies that $\mathscr{C}_{2,0}=\mathscr{C}_{2,0}(t;r_1,r_2)$, with $t=t_1-t_2$. 
Invariance under both non-local and local space-translations gives $Y_{-1}\mathscr{C}_{2,0}=Z_{-1}\mathscr{C}_{2,0}=0$. 
In Fourier space, this becomes 
\BD
\left(\vep_1 |q_1| +\vep_2 |q_2|\right)\wht{\mathscr{C}}_{2,0}(t;q_1,q_2)=0 \;\; , \;\; 
\left( q_1 + q_2\right)\wht{\mathscr{C}}_{2,0}(t;q_1,q_2)=0
\ED
where the signatures are both positive, viz. $\vep_1=\vep_2=+1$. The only solution is $\wht{\mathscr{C}}_{2,0}(t;q_1,q_2)=0$, as asserted. \hfill q.e.d.\\

Recall that the dynamical symmetry of the $\mathfrak{ev}$ algebra is only a symmetry of the deterministic part 
of the {\sc dle} Langevin equation (\ref{1}), which corresponds to
$T=0$. The vanishing of $\mathscr{C}_{2,0}$ is seen explicitly in the exact {\sc dle}-correlator (\ref{6C},\ref{7C}), which indeed vanishes as $T\to 0$.
This result of the {\sc dle} process is analogous to what is found for Schr\"odinger-invariant systems \cite{Henkel10}, 
where it follows from a Bargman superselection rule \cite{Bargman54}. 
\textcolor{black}{Still, this does not mean that symmetry methods could only predict vanishing correlators. 
For example, in Schr\"odinger-invariant systems, correlators
with $T\ne 0$ can be found from certain integrals of higher $n$-point responses \cite{Picone04,Henkel10}. 
For a simple illustration in the noisy Edwards-Wilkinson equation, see \cite{Henkel16c}. 
We conjecture that an analogous procedure might work for the {\sc dle} process and hope to return to this elsewhere.}

We now concentrate on the two-time {\it response function} $\mathscr{R}=\mathscr{R}(t_1,t_2;r_1,r_2)=\mathscr{C}_{1,1}(t_1,t_2;r_1,r_2)$. 
Time-translation-invariance, which imposes $X_{-1}\mathscr{R}=0$, implies that $\mathscr{R}=\mathscr{R}(t;r_1,r_2)$, with $t=t_1-t_2$. 
Invariance under non-local and local space-translations now give (in Fourier space)
\BD
\vep_1 \left(|q_1| -  |q_2|\right)\wht{\mathscr{R}}(t;q_1,q_2)=0 \;\; , \;\; 
\left( q_1 + q_2\right)\wht{\mathscr{R}}(t;q_1,q_2)=0
\ED
since the signatures are now $\vep_1=-\vep_2=+1$. 
Here, a non-vanishing solution is possible and we can write $\mathscr{R}=F(t,r)$, with $r=r_1-r_2$. 

\noindent
{\bf Proposition 6:}
{\it The $\mathfrak{ev}$-covariant two-point response function $\mathscr{R}=\mathscr{C}_{1,1}$ from (\ref{Cnm})
satisfies the scaling form $\mathscr{R}=\langle \vph_1(t,r)\wit{\vph}_2(0,0)\rangle=t^{-2x} f(v)$, with the scaling variable
$v=r/t$. If the scaling function $f(v)$ obeys the following two conditions, with the abbreviations $x=\demi(x_1+x_2)$ and $\xi=\demi(\xi_1+\xi_2)$,}
\BEQ \label{23}
\left( \vep_1 \nabla_v +\mu v\partial_v +2\mu\xi \right) f(v) = 0 \;\; , \;\;
(x_1-x_2) \left( \vep_1 \nabla_v +\mu v\partial_v + \mu \right) f(v) = 0 .
\EEQ
{\it and the constraint $\xi_1-\xi_2=x_1-x_2$ holds true, then all quasi-primary Ward identities are satisfied.}

\noindent The conditions (\ref{23}) come from the deterministic part of the {\sc dle} Langevin equation (\ref{1}) and do not contain $T$. 
This is consistent with the $T$-independence of the exact {\sc dle}-response function (\ref{6R},\ref{7R}).\footnote{A fuller justification,
analogous to derivation of the Bargman superselection rules of Schr\"odinger-invariance \cite{Henkel94,Picone04}, 
is left as an open problem, for future work.}  

\noindent
{\bf Proof:} 
Denote by $x_i$ and $\xi_i$ (with $i=1,2$), the two scaling dimensions of the scaling operator $\vph_1$ 
and of the response operator $\wit{\vph}_2$, respectively. 
Time-translation-invariance and non-local and local space-translation-invariances produced the form 
$\mathscr{R}=F(t,r)$, with $t=t_1-t_2$, $r=r_1-r_2$ and the signatures $\vep_1=-\vep_2=+1$. 
Using (\ref{16}), the other six Ward identities lead to 
\begin{subequations} \label{cov}
\begin{align}
\left[ -t\partial_t -r\partial_r -x_1-x_2 \right] F &= 0 \label{covX0} \\
\left[ -t\vep_1 \nabla_r -\mu r\partial_r -\mu\xi_1 - \mu\xi_2 \right] F &= 0 \label{covY0} \\
\left[ -t \partial_r -\mu \vep_1 r\nabla_r -\mu(\xi_1 + \xi_2)\vep_1 \partial_r \nabla_r^{-1} \right] F &= 0 \label{covZ0} \\
\left[ -t^2\partial_t -2tr\partial_r -\mu r^2 \vep_1 \nabla_r -2x_1 t -2\mu\xi_1 \vep_1 r \partial_r \nabla_r^{-1} \right] F &= 0 \label{covX1} \\
\left[ -t^2 \vep_1 \nabla_r -2\mu tr\partial_r -\mu^2 \vep_1 r^2 \nabla_r -2\mu\xi_1 t -2\mu^2\xi_1\vep_1 r\partial_r \nabla_r^{-1} \right] F &= 0
\label{covY1} \\
\left[ -t^2 \partial_r -2\mu \vep_1 tr\nabla_r -\mu^2 r^2 \partial_r -2\mu^2\xi_1 r -2\mu\xi_1\vep_1 t\partial_r\nabla_r^{-1}  \right] F &= 0
\label{covZ1} 
\end{align}
\end{subequations}
Herein, eq.~(\ref{covX1}) is obtained by using eqs.~(\ref{covX0},\ref{covZ0}), and eqs.~(\ref{covY1},\ref{covZ1}) 
are obtained by using (\ref{covY0},\ref{covZ0}). Actually, because of the identity 
\BEA
\left[ -t\partial_r -\mu \vep_1 r\nabla_r -2\mu\xi \vep_1 \partial_r\nabla_r^{-1}\right]F 
&=&  \vep_1 \left[ -t\vep_1 \nabla_r \partial_r \nabla_r^{-1} -\mu r \nabla_r^2 \nabla_r^{-1} -2\mu\xi \partial_r \nabla_r^{-1} \right] F 
\nonumber \\
&=& \vep_1 \partial_r \nabla_r^{-1} 
\left[ -\vep_1 t \nabla_r -\mu r \partial_r -2\mu\xi \right] F 
\nonumber 
\EEA
the condition $Y_0F=0$, eq.~(\ref{covY0}), implies $Z_0F=0$, eq.~(\ref{covZ0}). 
Since $\left(\partial_r\nabla_r^{-1}\right)^2 f(r)=f(r)$, see the corrollary~\ref{co1}, the converse also holds true. 
Next, eq.~(\ref{covX1}) can be simplified further:  multiply eq.~(\ref{covX0}) with $t$ and
subtract it from (\ref{covX1}), which gives
\BEQ \label{covX1simp}
\left[ -tr\partial_r - \mu\vep_1 r^2\nabla_r -(x_1-x_2)t -2\mu\xi_1\vep_1 r\partial_r\nabla_r^{-1}\right] F = 0
\EEQ 
Then multiply (\ref{covZ0}) with $r$ and substract it from (\ref{covX1simp}). This gives the condition
\BEQ \label{36}
\left[ \left( x_1 - x_2\right)t +\mu\left(\xi_1 - \xi_2\right)\vep_1 r \partial_r \nabla_r^{-1} \right] F = 0
\EEQ
Similarly, simplify eq.~(\ref{covY1}): multiply (\ref{covY0}) by $t$ and subtract from (\ref{covY1}), 
then multiply (\ref{covX1simp}) by $\mu$ and subtract as well. This gives
\BD
\mu \left[ \left(\xi_1 - \xi_2\right) - \left( x_1 - x_2\right) \right] t F = 0
\ED
Unless $F\sim \delta(t)$ is a distribution, this gives the constraint $\xi_1-\xi_2= x_1 - x_2$. 
Finally, eq.~(\ref{covZ1}) is simplified by multiplying
first (\ref{covZ0}) with $t$ and subtracting and then multiplying (\ref{covY0}) with $r$ and subtracting as well. This leads to
$(\xi_1 - \xi_2)\left[ \mu r +\vep_1 t \partial_r \nabla_r^{-1}\right]F=0$. 
Since in the proof of the corollary~\ref{co1}, we have seen that 
$\left( \partial_r \nabla_r^{-1}\right)^2 f(r) = f(r)$, this can be rewritten as follows:
$(\xi_1 - \xi_2)\vep_1 \left( \partial_r \nabla_r^{-1}\right) \left[ \vep_1 \mu r\partial_r\nabla_r^{-1} + t \right]F=0$. 
Taking the constraint into account, the last condition can be combined with (\ref{36}) into the single equation
\BEQ \label{37}
\left( x_1 - x_2\right) \left[ t +\mu\vep_1 r \partial_r \nabla_r^{-1} \right] F = 0
\EEQ
The form of $F$ is now fixed by the three equations (\ref{covX0},\ref{covY0},\ref{37}) and the constraint has to be obeyed. 

Eq.~(\ref{covX0}) implies the scaling form $F=t^{-2x} f(r/t)$. Inserting this into (\ref{covY0}) produces, with the help of (\ref{16}), the
first of the equations (\ref{23}). Finally, inserting the scaling form for $F$ into (\ref{37}) gives
$(x_1-x_2)\left( 1 + \mu\vep_1 v \partial_v \nabla_v^{-1}\right) f(v)=0$. 
Since it is not immediately obvious if that condition is consistent with the first eq.~(\ref{23}), we rephrase it as follows: 
use the commutator $\left[ v\partial_v, \nabla_v^{-1}\right] = \nabla_v^{-1}$ to write 
formally $v\partial_v \nabla_v^{-1} = \nabla_v^{-1} + \nabla_v^{-1} \left( v \partial_v\right)$. 
Then, apply $\nabla_v$ to the last condition on $f(v)$ derived from (\ref{37}), in order to rewrite it as follows
\BD
\nabla_v \nabla_v^{-1} \left[ \left( x_1 - x_2\right) \left( \vep_1 \nabla_v +\mu v \partial_v +\mu \right) \right] f(v) = 0
\ED
and this equation is obeyed if the second eq.~(\ref{23}) holds true. 
We have found a sufficient set of conditions to satisfy all nine {\sc dle} quasi-primary Ward identities
for $\mathscr{R}=\mathscr{C}_{1,1}$. \hfill q.e.d. \\

The two conditions in eq.~(\ref{23}) are compatible in two distinct cases:
\begin{enumerate}
\item[{\bf Case A}:] \textcolor{blue}{\underline{\textcolor{black}{$2\xi=1$}}}. 
Then $\left( \vep_1 \nabla_v +\mu v\partial_v + \mu \right) f(v) = 0$ and $x_1\ne x_2$ is still  possible. 
\item[~~~{\bf Case B}:] \textcolor{blue}{\underline{\textcolor{black}{$x_1=x_2$}}}. 
Then $\xi_1=\xi_2$ and $\left( \vep_1 \nabla_v +\mu v\partial_v +2\mu\xi \right) f(v) = 0$. 
\end{enumerate}
We must also compare the differential operator ${\cal S}=-\mu\partial_t+\nabla_r$ with the {\sc dle} Langevin equation (\ref{1}). 
Taking into account the normalisation in the definition of the Riesz-Feller derivative, we find $\mu^{-1} = \II \nu$. 
Physically, one should require $\nu>0$ in order that the correlators and responses vanish for large momenta $|q|\to\infty$. \\

\noindent  
{\bf Proposition 7:}
{\it The $\mathfrak{ev}$-co-variant two-time response function $\mathscr{R}_{12}(t,r)=F(t,r)$ has the form}
\BEA
F(t,r) &=&  F_A\, \delta_{\xi_1+\xi_2,1}\,\delta_{\xi_1-\xi_2,x_1-x_2}\; t^{1-x_1-x_2} \frac{\nu t}{\nu^2 t^2 + r^2} 
\nonumber \\
& & +  F_B\, \delta_{x_1,x_2}\,\delta_{\xi_1,\xi_2}\; t^{1+\psi-2x_1} \left( \nu^2 t^2 + r^2\right)^{-(\psi+1)/2} 
\cos\left( (\psi+1)\arctan\left(\frac{r}{\nu t}\right)-\frac{\pi\psi}{2}\right)~~~~~~
\label{Rfinal}
\EEA
{\it where $\psi = (\xi_1+\xi_2)-1$ is assumed real, 
$F_{A,B}$ are normalisation constants and the convention $\vep_1=+1$ is admitted.} 
 
\noindent
{\bf Proof:}
Both cases can be treated in the same way. The first eq.~(\ref{23}) becomes in Fourier space
\BD
\left[ \II\vep_1 |q|  -\mu q\partial_q +\mu(2\xi-1)\right] \wht{f}(q)=0
\ED
In case A, the constant term vanishes, while it is non-zero in case B. The solution reads 
\BD
\wht{f}(q) = \wht{f}_0\, q^{2\xi-1}\exp\left(\II\vep_1 |q|/\mu\right) = \wht{f}_0\, q^{2\xi-1}\exp\left(-\vep_1 \nu |q|\right) 
\ED
where $\wht{f}_0$ is a normalisation constant and we can now adopt $\vep_1=+1$. 
We also introduced the constant $\nu$ from the {\sc dle} Langevin equation (\ref{1}) to illustrate that $\wht{f}(q)\to 0$ for $|q|$ 
large when $\nu$ is positive. Both cases A and B produce valid solutions of the linear eqs.~(\ref{23}), such that the general solution should be a
linear superposition of both cases. The inverse Fourier transforms are carried out straightforwardly. \hfill q.e.d. \\

\noindent
{\bf Comment 3:}
Propositions~4 and~7 contain the assertions in the theorem, which are also listed in table~\ref{tab3}. 
Proposition~5 proves the statement in Comment~2. 
We had already mentioned in section~1 (Comment~1), that if we restrict to case A and take
$x=x_1=x_2=\demi$ and $\nu>0$, the resulting two-time response 
$F(t,r) = F_0\, t^{1-2x}\, {\vep_1 \nu t}/(\nu^2 t^2 + r^2)$, with $t=t_1-t_2$ and $r=r_1-r_2$,
reproduces the exact solution (\ref{7R}). 
We stress that no choice of $x_1$ will make the ortho-conformal prediction (\ref{Cortho}) compatible with (\ref{7R}). 

This is the main conceptual point of this work: {\em the non-local representation (\ref{metaconf_inf}) of the meta-conformal algebra 
$\mathfrak{ev}$ is necessary to reproduce the correct scaling behaviour of the non-stationary response of the {\sc dle} process.} 

The non-local meta-2 conformal invariance produces the response function $\mathscr{R}=\mathscr{C}_{1,1}$, 
whereas all local ortho-, galilean and meta-conformal invariances yielded a correlator $\mathscr{C}_{2,0}$. \\

\textcolor{black}{The main result (\ref{R_Thm},\ref{Rfinal}) on the shape of the meta-2-conformal response  can be cast into the scaling form 
$t^{x_1+x_2}\mathscr{R}_{12}(t,r)=f(r/t)$, with the explicit scaling function
\BEQ
f(u) = \frac{\left(1+u^2\right)^{-1} +\rho \left(1+u^2\right)^{-\xi_1} \sin\left(\pi\xi_1 -2\xi_1\arctan u\right)}{1+\rho \sin \pi\xi_1}.
\EEQ
We see that the first scaling dimensions $x_1,x_2$ merely arrange the data collapse, while the form of the scaling functions only depends
on the second scaling dimension $\xi_1=\xi_2$ and the amplitude ratio $\rho$ 
(the exact solution (\ref{7R}) of the {\sc dle}-process corresponds to $\rho=0$). 
The normalisation is chosen such that $f(0)=1$. For $\xi_1=\demi$,
we simply have $f(u)=(1+u^2)^{-1}$. In figure~\ref{fig4}, several examples of the shape of $f(u)$ are shown. Clearly, these are quite 
distinct from all the examples of ortho-, meta-1- and Galilean-conformal invariance, displayed above in figure~\ref{fig3}. 
\begin{figure}[tb]
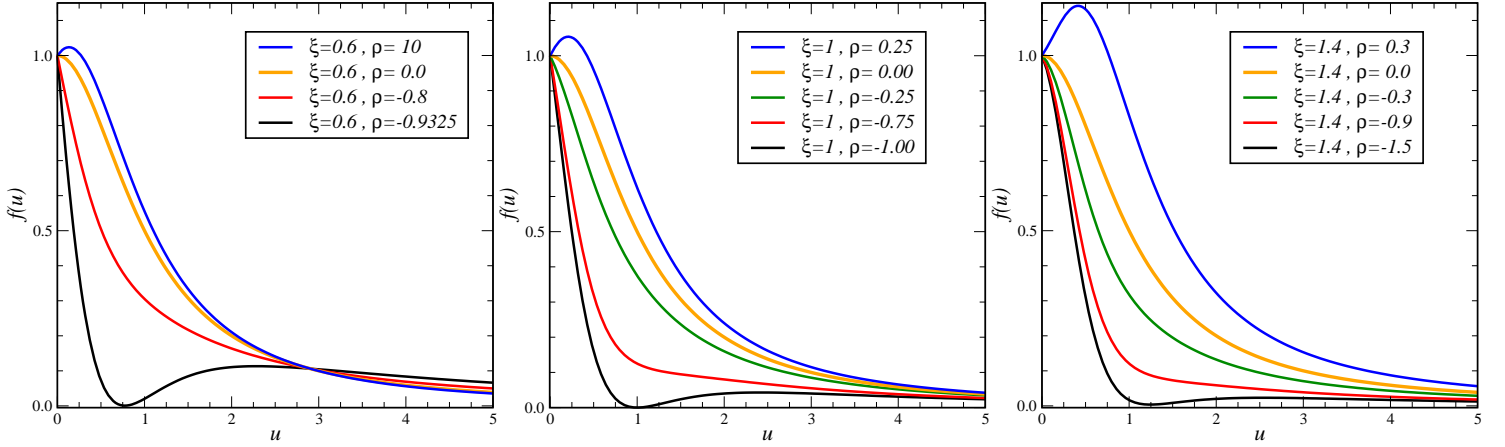

\centerline{\includegraphics[scale=0.465]{dle_xi06_scal}\hspace{-0.0truecm}\includegraphics[scale=0.465]{dle_xi10_scal}\hspace{-0.0truecm}\includegraphics[scale=0.465]{dle_xi14_scal}} 
\caption[fig4]{Scaling function $f(u)$ of the covariant meta-2-conformal two-point response $\mathscr{R}(t,r)=t^{-x_1-x_2}f(r/t)$, 
over against the scaling variable $u=r/t$, for $\xi=\xi_1=[0.6,1.0,1.4]$ in the left, middle and right panels, respectively, 
and several values of the amplitude ratio $\rho$. 
\label{fig4}}
\end{figure}
}

\noindent{\bf Comment 4:}
By analogy with Schr\"odinger-invariance, we expect that the fluctuation-dominated correlators should be 
obtained from certain time-space integrals of higher
$n$-point responses \cite{Picone04,Henkel10}. Working with the quantum chain representation of the terrace-step-kink model, 
Karevski and Sch\"utz have calculated the stationary two-point correlator of the densities, which in our terminology correspond 
to the slopes $u(t,r)=\partial_r h(t,r)$. They find \cite{Karevski16}
\BD
C(t,r) = C_A\, t^{-2}\: \frac{1-\xi^2}{(1+\xi^2)^2} + C_B\, t^{-\psi}\: \frac{\cos[2 (q^* r - \omega t)]}{(1+\xi^2)^{\psi}}
\ED
with the scaling variable $\xi=(r-v_c t)/(\nu t)$, where $v_c$ is the global velocity of the interface, $\psi\geq \demi$ 
is a real parameter and $C_{A,B}$ are known normalisation constants. 
The structure of their result is qualitatively very close to the form (\ref{R_Thm},\ref{Rfinal}) 
for the two-time response of the $\mathfrak{ev}$-algebra, in the sense that it contains
a dominant and monotonous term\footnote{It can be checked that from the exact height-height correlator (\ref{6C},\ref{7C}) 
this first term is recovered by computing
the correlator $\langle u(t,r) u(0,0)\rangle$ of the densities $u=\partial_r h$.} 
and a non-dominant and oscillatory one. We interpret this as an encouraging signal that
it should be possible to find the correlators from the $\mathfrak{ev}$ as well, by drawing on the analogies with Schr\"odinger-invariance. 
The first step in this direction would be the derivation of an analogue of a Bargman superselection rule, which is work in progress.\\

\noindent{\bf Comment 5:} 
\textcolor{black}{The consequences of the choice of the fractional derivative are difficult to appreciate in advance 
and largely remain a matter of try and error. 
Our choice of the Riesz-Feller derivative was suggested that in this way the Lie algebra becomes 
a dynamical symmetry of the {\sc dle} \textcolor{black}{process}. 
In the past, we had also worked \cite{Henkel02,Henkel10} with an extension of the Riemann-Liouville derivative by distributional terms \cite{Gelfand}. 
For dynamical exponents $z\ne 1,2$, this leads to a strong oscillatory behaviour
of the response functions which appears to be physically undesirable.\\
We consider the success of the simple case study of the {\sc dle} process} treated here as suggestive for future investigations.\\

\noindent
{\bf Corollary 4:} In the limit $\mu\to 0$ (or $\nu\to\infty$) the Lie algebra $\mathfrak{ev}$ can be contracted into the algebra
\BEQ \label{metacga_comm}
\left[ X_n, X_m \right] = (n-m) X_{n+m} \;\; , \;\; \left[ X_n, B_m^{\pm}\right] = (n-m) B_{n+m}^{\pm}
\EEQ
with $n,m\in\mathbb{Z}$ and with the explicit generators
\BEA
X_n &=& -t^{n+1}\partial_t -(n+1)t^n r\partial_r - (n+1) x t^n - n(n+1)\vep\gamma t^{n-1} r \partial_r \nabla_r^{-1} \nonumber \\
B_n^{\pm} &=& -\demi t^{n+1}\left( \vep \nabla_r \pm \partial_r \right) -\demi (n+1)\gamma t^n \left( 1 \pm \vep \partial_r \nabla_r^{-1}\right)
\label{metacga_inf}
\EEA
where $\vep=\pm 1$ is the signature and $x=x_{\vph}$ and $\gamma=\gamma_{\vph}$ 
are the scaling dimension and the rapidity of the scaling operator on which these generators
act. The co-variant quasi-primary two-point correlator $\mathscr{C}_{2,0}=0$, whereas the co-variant quasi-primary 
two-point response $\mathscr{R}=\mathscr{C}_{1,1}$ is, with the normalisation constant $\mathscr{R}_0$
\BEQ
\mathscr{R}(t,r) = \delta_{x_1,x_2} \delta_{\gamma_1,\gamma_2}\: \mathscr{R}_0\: t^{-2x_1} \exp\left(-2\left|\frac{\gamma_1 r}{t}\right| \right). 
\EEQ

The algebra (\ref{metacga_comm}), which one might call {\bf meta-conformal galilean algebra}, 
contains the conformal galilean algebra as a sub-algebra, although the
generators (\ref{metacga_inf}) are in general non-local, in contrast with those in eq.~(\ref{Tcga}). 
However, the co-variant two-point function is here a response, and {\em not} a correlator. 

\noindent 
{\bf Proof:}
In order to carry out the contraction on the generators (\ref{metaconf_inf}), where $X_n = A_n + B_n^+ + B_n^-$, we first change coordinates
$r\mapsto \mu r$, let $\xi_i=\gamma_i/\mu$ and rescale the generators $B_n^{\pm}\mapsto \mu B_n^{\pm}$. Then
the last commutator in (\ref{metaconf_ter}) becomes $\left[ B_n^{\pm}, B_m^{\pm}\right] = \mu (n-m) B_{n+m}^{\pm}$. 
Taking the limit $\mu\to 0$ produces the
generators (\ref{metacga_inf}) and the commutators (\ref{metacga_comm}) immediately follow. The Ward identities for the finite-dimensional 
sub-algebra are written down as before and $\mathscr{C}_{2,0}=0$ follows. 
For the response function, going again through the proof of the proposition~7 and 
recalling that $\mu^{-1}=\II\nu$, we see that case A in (\ref{Rfinal}) does not have a non-vanishing limit as $\nu\to\infty$. 
For case B, consider the scaling form $\mathscr{R}(t,r)=t^{-2x_1} f(r/t)$, with the scaling function $f(v)$ written as
\BD
f(v) = f_0 \left[ \left( \frac{\vep_1}{\II \mu} -i v\right)^{-\psi-1} + e^{\II\pi\psi}  \left( \frac{\vep_1}{\II \mu} +i v\right)^{-\psi-1} \right]
\ED
and $\psi+1=2\gamma_1/\mu$. If $\gamma_1>0$, the first term 
$\left( -\II/\mu -\II v\right)^{-\psi-1} = \left( \II\mu\right)^{2\gamma_1/\mu} \left( 1+\mu v\right)^{-2\gamma_1/\mu} \stackrel{\mu\to 0}{=}
\mu_0 e^{-2\gamma_1 v}$, and where $\mu_0$ is a constant, to be absorbed into the overall normalisation. 
The second term vanishes, since $e^{\II\pi\psi}=e^{\II\pi(-1+2\II\nu\gamma_1)}=- e^{-2\pi\nu\gamma_1}\to 0$ as $\nu\to\infty$. 
On the other hand, if $\gamma_1<0$, one divides $f(v)$ by $e^{\II\pi\psi}$, and redefines the normalisation constant. Now, the second term produces
$\sim e^{+2\gamma_1 v}$ and the first one vanishes in the $\nu\to\infty$ limit. Both cases are  combined into $f(v)=\bar{f}_0 e^{-2|\gamma_1 v|}$. 
Alternatively, one derives from the Ward identities the two constraints $x_1=x_2$ and $\gamma_1=\gamma_2$. Global dilation-invariance gives the
scaling form $\mathscr{R}(t,r) = t^{-2x_1} f(r/t)$ where the scaling function $f(v)$ must satisfy the equation
$f'(v) +2\gamma_1 \sign(t) f(v)=0$ which leads to the asserted form, modulo a dualisation procedure, 
analogous to \cite{Henkel15c,Henkel16b} to guarantee the boundedness for large separations. 
\hfill q.e.d. \\

Our results on non-local meta-conformal algebras are summarised in table~\ref{tab3}. \\

We close with a last example illustrating how local scale-invariance might help in the extraction of the values of non-equilibrium exponents. 

\begin{table}[tb]
\caption[tab3]{Comparison of non-local meta-2 conformal invariance, and meta-conformal galilei invariance in $(1+1)D$. 
The non-vanishing Lie algebra commutators, the defining equation of the generators and the invariant differential operator $\cal S$ 
are indicated. The usual generators are $X_n= A_n+B_n^{+}+B_n^{-}$, $Y_n=B_n^{+}+B_n^{-}$ and $Z_n=B_n^{+}-B_n^{-}$, see also table~\ref{tab2}. 
Physically, the co-variant quasiprimary two-point function $\mathscr{R}_{12}=\langle \vph_1(t,r)\wit{\vph}_2(0,0)\rangle$ is a response function. 
In case B, one has $\psi=2\xi_1-1$. \label{tab3}} 
\begin{center}\begin{tabular}{||c|ll|lr||} \hline\hline
                   & \multicolumn{1}{c}{meta-2 conformal~ }                     & \multicolumn{1}{c|}{meta-conformal galilean}         & \multicolumn{1}{c}{constraints} &   \\ \hline
Lie                & $\left[ A_n, A_m\right]~ = (n-m) A_{n+m}$                  & $\left[ X_n, X_m\right] = (n-m) X_{n+m}$             & & \\
algebra            & $\left[ B_n^{\pm}, B_m^{\pm}\right] \:=(n-m)B_{n+m}^{\pm}$ & $\left[ X_n, B_m^{\pm}\right] \:=(n-m)B_{n+m}^{\pm}$ & & \\[0.14truecm] \hline
generators         & ~~(\ref{metaconf_inf})                                     & ~~(\ref{metacga_inf})                                & & \\ \hline
${\cal S}$         & $-\mu\partial_t + \nabla_r$                                &                                                      & & \\[0.14truecm] \hline
                   & $t^{1-x_1-x_2}\cdot \nu t\left(\nu^2 t^2 + r^2\right)^{-1}$  &   --       & \multicolumn{2}{l||}{$\xi_1+\xi_2=1$  \hfill ~~~~({\bf A})} \\
                   &                                                            &              & \multicolumn{2}{l||}{$x_1-\xi_1=x_2-\xi_2$~ }  \\[0.24truecm]
$\mathscr{R}_{12}$ &  $t^{1-2x_1+\psi} \left( \nu^2 t^2 + r^2\right)^{-(\psi+1)/2}$ &  $t^{-2x_1}\exp\left(-2\left|\gamma_1 r/t\right|\right)$ & $x_1=x_2$ & ({\bf B}) \\
                   &  $\cdot \cos\left[\left(\psi+1\right)\arctan\left(\frac{r}{\nu t}\right) -\frac{\pi\psi}{2} \right]$ 
                   & & \multicolumn{2}{l||}{$\xi_1=\xi_2$ {\small , or $\gamma_1=\gamma_2$}~ }  \\[0.16truecm] \hline \hline     
\end{tabular}\end{center}
\end{table}

\noindent 
{\bf Example 6:}
The {\bf Kardar-Parisi-Zhang equation} \cite{Kardar86}
\BEQ 
\partial_t h(t,\vec{r}) = \nu \Delta_{\vec{r}}  h(t,\vec{r}) +\frac{\mu}{2} \left( \frac{\partial h(t,\vec{r})}{\partial \vec{r}}\right)^2  
+ \left( 2\nu T\right)^{1/2}\wht{\eta}(t,\vec{q})
\EEQ
where $\Delta_{\vec{r}}$ is the spatial laplacian and $\nu,\mu$ are constants, 
is a paradigmatic example of interface growth \cite{Barabasi95,Krug97}. Its autoresponse function
$R(t,s;\vec{0})$ and its autocorrelator $C(t,s;\vec{0})$ have been computed from large-scale numerical simulations 
and are seen to obey the scaling forms (\ref{4}), see table~\ref{tab1} for
the values of the exponents. In $d=1$ space dimensions, the system is integrable and many exact results are known. 
Concerning the non-stationary behaviour, we recall that a perturbative
renormalisation-group study predicts $\lambda_C=d$ for dimensions $d<2$ \cite{Krech97}, its domain of applicability. 
We also recall that by combining Galilei-invariance and
time-reversal invariance (the latter only holds for $d=1$) a fluctuation-dissipation relation ({\sc fdr}) can be found, in the stationary limit, 
where $t,s\to\infty$ such that $\tau =t-s$ is kept fixed, 
which reads $T R(\tau;r)=-\partial_r^2 C(\tau;r)$ \cite{Lvov93,Frey96,Henkel12}.\footnote{This is distinct from Kubo's  {\sc fdr}
$T R(\tau;r)=-\partial_{\tau} C(\tau;r)$ which holds at equilibrium.}
Combination with the scaling ansatz (\ref{4}) gives the exponent relations 
\BEQ \label{exposants}
\lambda_C=\lambda_R \;\; \mbox{\rm ~and~} \;\; 1+a=b+2/z
\EEQ
which are satisfied by the exact values of table~\ref{tab1}, for $d=1$. 
On the other hand, analysing the available high-quality numerical data for $d=2$ \cite{Odor14,Halp14} 
with the simple scaling ansatz (\ref{4}) has cast doubts on the validity of both relations 
$\lambda_C\stackrel{?}{=}2$ and $\lambda_C\stackrel{?}{=}\lambda_R$. 

\noindent Very recently, it has been shown that taking into account the proper dynamical symmetries can modify this perception. 
New information can be obtained on the scaling of the autoreponse function $R(t,s;\vec{0})$, by borrowing techniques from 
`logarithmic conformal invariance'. Formally, this amounts to replace scaling dimensions by (non-diagonalisable) Jordan matrices. 
Since the scaling form of the autoresponse function only contains the exponent $\lambda_R/z$, the value of $z$ does not enter independently. 
Adapting logarithmic conformal theory to non-equilibrium dynamics, and taking into account that time-translations are not present, one arrives
at the prediction \cite{Henkel10b} (in the limit $s\to\infty$ and $y>1$ fixed) 
\BEQ \label{43}
R(ys,s;\vec{0}) = s^{-1-a} y^{-\lambda_R/z} \left( 1 - \frac{1}{y}\right)^{-1-a'} 
\left[ h_0 - g_0 \ln\left(1-\frac{1}{y}\right) -\frac{f_0}{2}\ln^2\left(1-\frac{1}{y}\right)\right]
\EEQ
with the three independent exponents $a,a',\lambda_R/z$ and the normalisation constants $f_0,g_0,h_0$.  

\noindent
For $d=1$, eq.~(\ref{43}) is in excellent agreement with the numerical data \cite{Henkel12} 
and the two exponent relations $\lambda_R=\lambda_C=1$ are obeyed. 
Remarkably, a recent thorough study \cite{Kelling16a,Kelling16b}
also produces this agreement in $d=2$ dimensions, taking into account the precise form (\ref{43}) of the scaling function.  
This gives good numerical evidence \cite{Kelling16b,Kelling16c} that both 
relationships $\lambda_R=\lambda_C=2$ should be satisfied, see table~\ref{tab1}. 
The second relation reproduces the prediction from \cite{Krech97}.  
{\it Is the evidence for the first eq.~(\ref{exposants}) a hint towards the existence of a yet unknown fluctuation-dissipation relation 
between $C$ and $R$, for $d=2$~?} \\

Hence a more precise knowledge on the expected scaling function can help to find the values of the exponents more precisely, from simulational data. \\

{\small
\noindent {\bf Acknowledgments:}
I thank X. Durang, J. Kelling and G. \'Odor for useful discussions or correspondence.  
This work was partly supported by the Coll\`ege Doctoral Nancy-Leipzig-Coventry
(Syst\`emes complexes \`a l'\'equilibre et hors \'equilibre) of UFA-DFH. 
}

{\small                      

}               


\begin{thebibliography}{999} 

\bibitem{Aarao07} 
  F.D.A. Aar\~ao Reis, J. Stafiej, Crossover of interface growth dynamics during corrosion and passivation,
  {\rm J. Phys. Cond. Matt.} {\bf 19}, 065125 (2007) {\tt [cond-mat/0609428]}. 
\bibitem{Bagchi09b} A. Bagchi, R. Gopakumar, I. Mandal, A. Miwa, GCA in $2D$, {\em JHEP}, 1008:004 (2010) {\tt [arxiv:0912.1090]}.
\bibitem{Barabasi95} 
  A.-L. Barab\'asi and H.E. Stanley H.E., {\em Fractal Concepts in  Surface Growth}, Cambridge University Press (Cambridge 1995). 
\bibitem{Bargman54} 
  V. Bargman, Unitary ray representations of continuous groups, {\rm Ann. of Math.} {\bf 56}, 1 (1954).
\bibitem{Barnich07} 
  G. Barnich, G. Comp\`ere, Classical central extension for asymptotic symmetries at null infinity in three spacetime dimensions,
  {\rm Class. Quant. Grav.} {\bf 24} F15 (2007) {\tt [arxiv:gr-qc/0610130]}.
\bibitem{Barnich07b} 
  G. Barnich, G. Comp\`ere, Classical central extension for asymptotic symmetries at null infinity in three spacetime dimensions (corrigendum), 
 {\rm Class. Quant. Grav.} {\bf 24}, 3139 (2007).
\bibitem{Belavin84} 
  A.A. Belavin, A.M. Polyakov, A.B. Zamolodchikov, 
  Infinite conformal symmetry in two-dimensional quantum field-theory, {\rm Nucl. Phys.} {\bf B241}, 333 (1984). 
\bibitem{Bert15} 
   L. Bertini, A. De Sole, D. Gabrielli, G. Jona-Lasinio, C. Landim, Macroscopic fluctuation theory, 
   {\rm Rev. Mod. Phys.} {\bf 87}, 593 (2015) {\tt [arXiv:1404.6466]}. 
\bibitem{Cartan1909} 
  \'E. Cartan, Les groupes de transformation continus, infinis, simples, 
  {\rm Ann. \'Ecole Norm. 3e s\'erie} {\bf 26}, 93 (1909).  
\bibitem{Cherniha10} 
  R. Cherniha, M. Henkel, The exotic conformal Galilei algebra and non-linear partial differential equations, 
  {\rm J. Math. Anal. Appl.} {\bf 369}, 120 (2010) {\tt [arxiv:0910.4822]}. 
\bibitem{CherMastKriv16}
  D. Chernyavsky, Coest spaces and Einstein manifolds with $\ell$-conformam Galilei symmetry, 
  Nucl. Phys. {\bf B911}, 474 (2016) {\tt [arxiv:1606.08224]};\\
  I. Masterov, Remark on  higher-derivative mechanics with $\ell$-conformal Galilei symmetry, 
  {\rm J. Math. Phys.} {\bf 57}, 092901 (2016) {\tt [arxiv:1607.02693]}; \\
  S. Krivonos, O. Lechtenfeld, A. Sorin, Minimal realization of $\ell$-conformal Galilei algebra, 
      Pais-Uhlenbeck oscillators and their deformation, JHEP {\bf 1610}, 073 (2016) {\tt [arxiv:1607.03756]};\\
  D. Chernyasky, A. Galajinsky, Ricci-flat space-times with $\ell$-conformal Galilei symmetry, 
  {\rm Phys. Lett.} {\bf B754},  249 (2016) {\tt [arxiv:1512.06226]};\\
  K. Andrezejewski, A. Galajinsky, J. Gonera, I. Masterov, Conformal Newton-Hooke symmetry of Pais-Uhlenbeck oscillator, 
  {\rm Nucl. Phys.} {\bf B885}, 150 (2014) {\tt [arxiv:1402.1297]};\\
  A. Galajinsky, I. Masterov, Dynamical realisation of $\ell$-conformal Newton Hooke group, 
  {\rm Phys. Lett.} {\bf B723}, 190  (2013) {\tt [arxiv:1303.3419]}. 
\bibitem{Cinti15} 
  E. Cinti, F. Ferrari,  Geometric inequalities for fractional Laplace operators and applications, 
  {\rm Nonlinear Differ. Equ. Appl.} {\bf 22}, 1699 (2015).
\bibitem{Duval09} 
  C. Duval, P.A. Horv\'athy, Non-relativistic conformal symmetries and Newton-Cartan structures, 
  {\rm J. Phys. A: Math. Theor.} {\bf 42}, 465206 (2009) {\tt [arxiv:0904.0531]}. 
\bibitem{Ebbinghaus08}
  M. Ebbinghaus, H. Grandclaude, M. Henkel,  Absence of logarithmic scaling in the ageing behaviour of the $4D$ spherical model,
  {\rm Eur. Phys. J.} {\bf B63}, 85 (2008) {\tt [arxiv:0709.3220]}. 
\bibitem{Edwards82} 
  S.F. Edwards, D.R. Wilkinson, The surface statistics of a granular aggregate, {\rm Proc. Roy. Soc.},  {\bf A381}, 17 (1982). 
\bibitem{Family85} 
  F. Family, T. Vicsek, Scaling of the active zone in the Eden process on percolation networks and the ballistic deposition model, 
  {\rm J. Phys. A Math. Gen.}, {\bf 18}, L75 (1985).
\bibitem{Francesco97} 
  P. di Francesco, P. Mathieu, D. S\'en\'echal,  {\em Conformal field-theory}, Springer(Heidelberg 1997). 
\bibitem{Frey96}
  E. Frey, U.C. T\"auber, T. Hwa, Mode-coupling and renormalization-group results for the noisy Burgers equation, 
  {\rm Phys. Rev.} {\bf E53}, 4424 (1996)  {\tt [arxiv:cond-mat/9601049]}. 
\bibitem{Gelfand} 
  I.M. Gel'fand and G.E. Shilov, {\it Generalized functions, vol. 1: properties and operations}, Academic Press (New York 1964). 
\bibitem{Hase06}
  M.O. Hase, S.R. Salinas, Dynamics of a mean spherical model with competing interactions, 
  {\rm J. Phys. A: Math. Gen.} {\bf 39}, 4875 (2006) {\tt [arxiv:cond-mat/0512286]}. 
\bibitem{Halpin95}
  T. Halpin-Healy, Y.-C. Zhang, Kinetic roughening phenomena, stochastic growth, directed polymers and all that, 
  {\rm Phys. Rep.} {\bf 254}, 215 (1995). 
\bibitem{Halp14} 
  T. Halpin-Healy, G. Palansantzas, Universal correlators and distributions as experimental signatures of 2+1 Kardar-Parisi-Zhang growth, 
  {\rm Europhys. Lett.} {\bf 105}, 50001 (2014) {\tt [arxiv:1403.7509]}. 
\bibitem{Havas78} 
  P. Havas, J. Plebanski, Conformal extensions of the Galilei group and their relation to the Schr{\"o}dinger group, 
  {\rm J. Math. Phys.},  {\bf 19}, 482 (1978). 
\bibitem{Henkel94} 
  M. Henkel, Schr\"odinger-invariance and strongly anisotropic critical systems, 
  {\rm J. Stat. Phys.} {\bf 75}, 1023 (1994) {\tt [arxiv:hep-th/9310081]}.
\bibitem{Henkel97} 
  M. Henkel, Extended scale-invariance in strongly anisotropic equilibrium critical systems, 
  {\rm Phys. Rev. Lett.} {\bf 78}, 1940  (1997) {\tt [arxiv:cond-mat/9610174]}.
\bibitem{Henkel02} 
  M. Henkel, Phenomenology of local scale invariance: from conformal invariance to dynamical scaling, 
  {\rm Nucl. Phys.}  {\bf B641}[FS], 405 (2002) {\tt [arxiv:hep-th/0205256]}. 
\bibitem{Henkel06b} M. Henkel, R. Schott, S. Stoimenov, J. Unterberger, 
  The Poincar\'e algebra in the context of ageing systems: Lie structure, representations, Appell systems and coherent states, 
  {\rm Confluentes Mathematici} {\em 4} 1250006 (2012), {\tt [arxiv:math-ph/0601028]}.  
\bibitem{Henkel10} 
  M. Henkel and M. Pleimling, {\it Non-equilibrium phase transitions vol. 2: 
  ageing and dynamical scaling far from equilibrium}, Springer (Heidelberg 2010). 
\bibitem{Henkel10b} 
  M. Henkel,  On logarithmic extensions of local scale-invariance, 
  {\rm Nucl. Phys.},  {\bf B869}[FS], 282 (2013) {\tt [arxiv:1009.4139]}
\bibitem{Henkel12} 
  M. Henkel, J.D. Noh, M. Pleimling, Phenomenology of ageing in the Kardar-Parisi-Zhang equation, 
  {\rm Phys. Rev.} {\bf E85}, 030102(R) (2012) {\tt [arxiv:1109.5022]}. 
\bibitem{Henkel15}
  M. Henkel, S. Stoimenov, Physical ageing and Lie algebras of local scale-invariance, In 
  {\it Lie Theory and its Applications in Physics}, V. Dobrev Ed., 
  Springer Proc. Math. Stat. Vol. 111, Springer (Heidelberg 2015); pp. 33-50; {\tt [arxiv:1401.6086]}. 
\bibitem{Henkel15b} 
  M. Henkel, X. Durang,  Spherical model of interface growth, 
  {\rm J. Stat. Mech.} {\bf },  P05022 (2015) {\tt [arxiv:1501.07745]}.  
\bibitem{Henkel15c} 
  M. Henkel, Dynamical symmetries and causality in non-equilibrium phase transitions, 
  {\rm Symmetry} {\bf 7}, 2108 (2015) {\tt [arxiv:1509.03669]}.
\bibitem{Henkel16a} 
  M. Henkel, Non-local meta-conformal invariance in diffusion-limited erosion, 
  {\rm J. Phys. A Math. Theor.} {\bf 49}, 49LT02 (2016) {\tt [arXiv:1606.06207]}. 
\bibitem{Henkel16b} 
  M. Henkel, S. Stoimenov, Meta-conformal invariance and the boundedness of two-point correlation functions, 
  {\rm J. Phys. A Math. Theor.} {\bf 49}, 47LT01 (2016) {\tt [arXiv:1607.00685]}. 
\bibitem{Henkel16c} 
  M. Henkel, From dynamical scaling to local scale-invariance: a tutorial,
  {\rm Eur. Phys. J. Spec. Topics}, (2017) to be published, {\tt [arxiv:1610.06122]}. 
\bibitem{Hosseiny10b}
  A. Hosseiny, S. Rouhani, Affine extension of galilean conformal algebra in $2+1$ dimensions,
  {\rm J. Math. Phys.} {\bf 51}, 052307 (2016) {\tt [arXiv:0909.1203]}.
\bibitem{Jacobi1843} 
  C.G. Jacobi, {\it Vorlesungen {\"u}ber Dynamik (1842/43), 4. Vorlesung}, in {\it ``Gesammelte Werke''}, 
  A. Clebsch and E. Lottner (eds), Akademie der Wissenschaften (Berlin 1866/1884) 
\bibitem{Kardar86} 
  M. Kardar, G. Parisi, Y.-C. Zhang, Dynamic scaling of growing interfaces, {\rm Phys. Rev. Lett.}, {\bf 56}, 889 (1986). 
\bibitem{Karevski16} 
   D. Karevski, G.M. Sch\"utz, Conformal invariance in driven diffusive systems at high currents, {\tt [arxiv:1606.04248]}. 
\bibitem{Kelling16a} 
   J. Kelling, G. \'Odor, S. Gemming,  Universality of $(2+1)$-dimensional restricted solid-on-solid models, 
   {\rm Phys. Rev.} {\bf E94}, 022107 (2016) {\tt [arxiv:1605.02620]}. 
\bibitem{Kelling16b} 
   J. Kelling, G. \'Odor, S. Gemming, Local scale-invariance of the $(2+1)$-dimensional Kardar-Parisi-Zhang model, 
   {\tt [arxiv:1609.05795]}. 
\bibitem{Kelling16c} G. Kelling, private communication (2016). 
\bibitem{Kloss12} 
   T. Kloss, L. Canet, N. Wschebor, Nonperturbative renormalization group for the stationary Kardar-Parisi-Zhang equation: 
   scaling functions and amplitude ratios in $1+1$, $2+1$ and $3+1$ dimensions,
   {\rm Phys. Rev.} {\bf E86}, 051124 (2012) {\tt [arxiv:1209.4650]}. 
\bibitem{Krug81} 
  J. Krug, P. Meakin, Kinetic roughening of laplacian fronts, {\rm Phys. Rev. Lett.} {\bf 66}, 703  (1991). 
\bibitem{Krug94} 
  J. Krug, Statistical physics of growth processes,  in  {\it Scale invariance, interfaces and non-equilibrium dynamics}, 
  A. McKane {\it et al.} (eds), 
  NATO ASI Series vol. {\em B344}, Plenum Press (London 1994), p. 1. 
\bibitem{Krug97} J. Krug, Origins of scale-invariance in growth processes, {\rm Adv. Phys.} {\bf 46}, 139 (1997). 
\bibitem{Krech97} 
  M. Krech, Short-time scaling behavior of growing interfaces, {\rm Phys. Rev.} {\bf E55}, 668 (1997) {\tt [cond-mat/9609230]}; \\ 
  erratum {\rm Phys. Rev.} {\bf E56}, 1285 (1997). 
\bibitem{Lie1881} S. Lie,  
  \"Uber die Integration durch bestimmte Integrale von einer Klasse linearer partieller Differentialgleichungen,
  {\rm Arch. for Mathematik og Naturvidenskab} {\bf 6}, 328 (1881). 
\bibitem{Lvov93}
  V.S. L'vov, V.V. Lebedev, M. Paton, I. Procaccia, Proof of scale invariant solutions in the Kardar-Parisi-Zhang and Kuramoto-Sivashinsky equations 
  in 1+1 dimensions: analytical and numerical results, {\rm Nonlinearity} {\bf 6}, 25 (1993). 
\bibitem{Negro97} 
  J. Negro, M.A. del Olmo, A. Rodr\'{\i}guez-Marco, 
  Nonrelativistic conformal groups, {\rm J. Math. Phys.} {\bf 38}, 3786 (1997).
\bibitem{Negro97b} 
  J. Negro, M.A. del Olmo, A. Rodr\'{\i}guez-Marco, 
  Nonrelativistic conformal groups II, {\rm J. Math. Phys.} {\bf 38}, 3810 (1997). 
\bibitem{Nezza12} 
  E. di Nezza, G. Palatucci, E. Valdinoci,  Hitchhiker's guide to the fractional Sobolev spaces, 
  {\rm Bull. Sci. Math\'ematiques} {\bf 136}, 521 (2012). 
\bibitem{Niederer72} 
  U. Niederer, The maximal kinematical invariance group of the free Schr\"odinger equation, 
  {\rm Helv. Phys. Acta} {\bf 45}, 802 (1972).
\bibitem{Odor14} 
  G. \'Odor, J. Kelling, S. Gemming, Ageing of the $(2+1)$-dimensional Kardar-Parisi-Zhang model, 
  {\rm Phys. Rev.} {\bf E89}, 032146 (2014) {\tt [arxiv:1312.6029]}. 
\bibitem{Ovsienko98} 
  V. Ovsienko, C. Roger, Generalisations of Virasoro group and Virasoro algebras through extensions by modules of tensor-densities on $S^1$, 
  {\rm Indag. Math.} {\bf 9}, 277 (1998). 
\bibitem{Paulos16} 
  M.F. Paulos, S. Rychkov, B.C. van Rees, B. Zan, Conformal Invariance in the Long-Range Ising Model,
  {\rm Nucl. Phys.} {\bf B902}, 249 (2016) {\tt [arxiv:1509.00008]}. 
\bibitem{Picone04} 
  A. Picone, M. Henkel,  Local scale-invariance and ageing in noisy systems, 
  {\rm Nucl. Phys.}  {\bf B688} 217, (2004) {\tt [arxiv:cond-mat/0402196]}.
\bibitem{Polyakov70} 
  A.M. Polyakov, Conformal symmetry of critical fluctuations, {\rm Sov. Phys. JETP Lett.} {\bf 12}, 381 (1970). 
\bibitem{Popk11} 
  V. Popkov, G.M. Sch\"utz, Transition probabilities and dynamic structure factor in the ASEP conditioned on strong flux, 
  {\rm J. Stat. Phys.} {\bf 142}, 627 (2011) {\tt [arxiv:1011.3913]}. 
\bibitem{Rodr15} 
  E.A. Rodrigues, B.A. Mello, F.A. Oliveira, Growth exponents of the etching model in high dimensions, 
  {\rm J. Phys. A Math. Theor.}  {\bf 48}, 035001 (2015).
\bibitem{Rodrigues15b} 
  E.A. Rodrigues, F.A. Oliveira, B.A. Mello, On the existence of an upper critical dimension for systems within the KPZ universality class, 
  {\rm Acta. Phys. Pol.} {\bf B46}, 1231 (2015)  {\tt [arxiv:1502.06121]}. 
\bibitem{Roet06} 
  A. R\"othlein, F. Baumann, M. Pleimling, Symmetry-based determination of space-time functions in nonequilibrium growth processes,
  {\rm Phys. Rev.} {\bf E74}, 061604 (2006), {\tt [arxiv:cond-mat/0609707]}; \\
  erratum  {\rm Phys. Rev.} {\bf E76}, 019901(E) (2007). 
\bibitem{Samk93} S.G. Samko, A.A. Kilbas, O.I. Marichev O.I.; {\it Fractional integrals and derivatives}, Gordon and Breach (Amsterdam 1993). 
\bibitem{Sethuraman16} 
  S. Sethuraman, On microscopic derivation of a fractional stochastic Burgers equation, 
  {\rm Comm. Math. Phys.} {\bf 341}, 625 (2016) {\tt [arxiv:1409.0944]}. 
\bibitem{Spohn99} 
  H. Spohn, Bosonization, vicinal surfaces, and hydrodynamic fluctuation theory, 
  {\rm Phys.  Rev.} {\bf E60}, 6411 (1999) {\tt [arxiv:cond-mat/9908381]}. 
\bibitem{Spohn14} 
  H. Spohn, Nonlinear fluctuating hydrodynamics for anharmonic chains, {\rm J. Stat. Phys.} {\bf 154}, 1191 (2014) {\tt [arXiv:1305.6412]}.  
\bibitem{Stoimenov15} 
  S. Stoimenov, M. Henkel, From conformal invariance towards dynamical symmetries of the collisionless Boltzmann equation, 
  {\rm Symmetry} {\bf  7}, 1595 (2015) {\tt [arxiv:1509.00434]}. 
\bibitem{Taeuber14} 
  U.C. T\"auber, {\em Critical dynamics: a field theory approach to equilibrium and non-equilibrium scaling behaviour},
  Cambridge University Press (Cambridge 2014).
\bibitem{Yeun96} 
  C. Yeung, M. Rao, R.C. Desai, Bounds on the decay of the auto-correlation in phase ordering dynamics, 
 {\rm Phys. Rev.} {\bf E53}, 3073 (1996) {\tt [cond-mat/9409108]}.
\bibitem{Yoon03} 
  S.Y. Yoon, Y. Kim, Surface growth models with a random-walk-like nonlocality, {\rm Phys. Rev.} {\bf E68}, 036121 (2003). 
\bibitem{Zhang10} 
  P.-M. Zhang,  P.A. Horv\'athy, Non-relativistic conformal symmetries in fluid mechanics,
  {\rm Eur. Phys. J.} {\bf C65}, 607 (2010)  {\tt [arxiv:0906.3594]}. 
\bibitem{Zoia07} 
  A. Zoia, A. Rosso, M. Kardar, Fractional Laplacian in Bounded Domains, 
  {\rm Phys. Rev.} {\bf E76}, 021116 (2007) {\tt [arxiv:0706.1254]}. 





\end{thebibliography}
\end{document}